\documentclass[11pt,a4paper]{article}
\usepackage{a4wide}

\usepackage[latin9]{luainputenc}
\usepackage{color}
\usepackage{mathrsfs}
\usepackage{amsmath}
\usepackage{amsthm}
\usepackage{amssymb}
\usepackage{esint}
\usepackage[unicode=true,pdfusetitle,
 bookmarks=true,bookmarksnumbered=false,bookmarksopen=false,
 breaklinks=false,pdfborder={0 0 1},backref=false,colorlinks=true]
 {hyperref}
\usepackage{cleveref}

\makeatletter

\@ifundefined{pageheight}{\let\pageheight\pdfpageheight}{}
\@ifundefined{pagewidth}{\let\pagewidth\pdfpagewidth}{}
\pageheight\paperheight
\pagewidth\paperwidth

\theoremstyle{plain}
\newtheorem{thm}{\protect\theoremname}
  \theoremstyle{definition}
  \newtheorem{defn}[thm]{\protect\definitionname}
  \theoremstyle{remark}
  \newtheorem{rem}[thm]{\protect\remarkname}
  \theoremstyle{remark}
  \newtheorem{claim}[thm]{\protect\claimname}
  \theoremstyle{plain}
  \newtheorem{assumption}[thm]{\protect\assumptionname}
  \theoremstyle{definition}
  \newtheorem{example}[thm]{\protect\examplename}
  \theoremstyle{plain}
  \newtheorem{cor}[thm]{\protect\corollaryname}
  \theoremstyle{remark}
  \newtheorem*{claim*}{\protect\claimname}
 \theoremstyle{definition}
 \newtheorem*{defn*}{\protect\definitionname}
  \theoremstyle{plain}
  \newtheorem*{cor*}{\protect\corollaryname}

\usepackage{marvosym}
\usepackage{hyperref}
\usepackage{bm}
\usepackage{xcolor}
\usepackage{enumerate}

\usepackage{color}

\usepackage{graphicx}
\usepackage{epstopdf}

\usepackage{soul}

\usepackage[normalem]{ulem}

\usepackage{subcaption}

\newcommand{\dif}{\mathrm{d}}

\renewcommand{\Re}{\operatorname{Re}}

\newcommand{\comment}[1]{}

\author{Jacob Shapiro
\\
\small{Theoretische Physik, ETH Z\"urich, 8093 Z\"urich, Switzerland} }

\usepackage{dsfont}
\usepackage{braket}
\let\oldnorm\norm
\def\norm{\@ifstar{\oldnorm}{\oldnorm*}}
\makeatother

\usepackage{tikz}
\usetikzlibrary{arrows}
\usepackage{pgfplots}
\usepackage{amsmath}
\usepackage{physics}
\usepackage{amssymb}

\makeatother

  \providecommand{\assumptionname}{Assumption}
  \providecommand{\claimname}{Claim}
  \providecommand{\corollaryname}{Corollary}
  \providecommand{\definitionname}{Definition}
  \providecommand{\examplename}{Example}
  \providecommand{\remarkname}{Remark}
\providecommand{\theoremname}{Theorem}

\begin{document}

\title{The Bulk-Edge Correspondence in Three Simple Cases}

\date{\today}
\maketitle
\begin{abstract}
We present examples in three symmetry classes of topological insulators
in one or two dimensions where the proof of the bulk-edge correspondence
is particularly simple. This serves to illustrate the mechanism behind
the bulk-edge principle without the overhead of the more general proofs
which are available. We also give a new formula for the $\mathbb{Z}_{2}$-index
of our time-reversal invariant systems inspired by Moore and Balents.
\end{abstract}

\section{Introduction}

Before the advent of topological insulators \cite{Hasan_Kane_2010}
and the Kitaev table \cite{Kitaev2009} the Bulk-Edge correspondence
was studied already in the context of the integer quantum Hall effect
(IQHE henceforth). Here the topological invariant of the system is
the integer Hall conductance, which can be viewed either as current
from the skipping electron orbits along the edge \emph{or} as the
drift of the cyclotron orbits within the bulk of the system. In an
actual physical sample the two mechanisms are at play according to
the capture potential at the boundaries. That the two integer numbers
should be equal is not a-prior obvious yet it has been proven already
by Hatsugai \cite{Hatsugai_PhysRevLett.71.3697} for the Harper model
via transfer matrix and Riemann-sheets. Later more general proofs
emerged (\cite{SBKR_2000}, \cite{Elbau_Graf_2002}, \cite{EGS_2005})
but at the greater cost of introducing very heavy machinery from C-star
algebra K-Theory or functional analysis. Meanwhile after the stellar
discoveries of \cite{Kane_Mele_2005_PhysRevLett.95.146802} and then
\cite{Schnyder_Ryu_Furusaki_Ludwig_PhysRevB.78.195125} it became
clear that there are additional interesting cases of topology other
than the IQHE, each carrying a topological invariant (in general $\mathbb{Z}$
or $\mathbb{Z}/2\mathbb{Z}$ valued) and for each a bulk and analogous
edge description available, with more general bulk-edge correspondence
proofs (\cite{PSB_2016}, \cite{Kubota2016} and \cite{Bourne2017}).

In view of all these developments the present paper has the modest
goal of asking if in very simple special cases one could understand
the principle behind the bulk-edge correspondence in more basic terms.
The answer is yes and three case studies are presented: the two-band
IQHE, the two-band chiral one-dimensional topological insulator, and
the four-band time-reversal-invariant topological insulator. Our aim
was to find the simplest yet non-trivial cases. Simplicity meant tight-binding
models on the lattice, translation invariance, nearest-neighbor hopping
and a minimum number of bands: two unless symmetry dictated otherwise
as in the case of time-reversal-invariance via Kramer's degenracy.
Sometimes we make even further simplifying assumptions which however
don't mean the system becomes trivial. Actually \cite{Mong_Shivamoggi_PhysRevB.83.125109}
already made the first step in this direction with an analysis of
Dirac Hamiltonians, and our work builds on theirs except for our first
IQHE proof which is independent.

The paper is organized as follows: We start by making precise the
simplifications we assume on our models, defining the bulk and edge
Hamiltonians and giving a schematic explanation of what the bulk-edge
correspondence means. In \cref{sec:The-1D-Chiral} we present the first
example in this work, which however relies on \cref{subsec:Dirac-Hamiltonians},
of the chiral one-dimensional case. We go on to \cref{sec:The-Integer-Quantum-Hall}
where we give two proofs for our simplified IQHE: the first one is
independent of \cref{subsec:Dirac-Hamiltonians} and is perhaps the
simplest of the collection, the second is essentially the proof of
\cite{Mong_Shivamoggi_PhysRevB.83.125109} reproduced here. Finally
we turn to odd time-reversal invariant systems in \cref{sec:The-2D-Time-Reversal}
where we provide a slight-generalization of the formula for the $\mathbb{Z}_{2}$-index
for inversion-symmetric systems given in \cite{Fu_Kane_2007_PhysRevB.76.045302}
and continue with another proof of its validity using a geometric
picture inspired by \cite{Moore_Balents_2007_PhysRevB.75.121306}.
We finish with another proof of the bulk-edge correspondence in this
case, which also relies on \cref{subsec:Dirac-Hamiltonians}.

A few comments are in order. We included here \cref{subsec:Dirac-Hamiltonians}
which already appeared in \cite{Mong_Shivamoggi_PhysRevB.83.125109}
in the interest of making the presentation self-contained. Since our
main focus here is to find simple manifestations of the bulk-edge
correspondence rather than make a review, we omitted many proofs (in
particular that the invariants are well-defined) and details about
the background of their definition which can easily be found in the
literature, starting from \cite{Hasan_Kane_2010}.

\section{\label{sec:Setting}Setting}

We work in the single-particle, tight-binding approximation so that
the underlying Hilbert space is either $\mathcal{H}=l^{2}\bigl(\mathbb{Z}^{d}\bigr)\otimes\mathbb{C}^{N}$
for the bulk or $\mathcal{H}^{\sharp}=l^{2}\bigl(\mathbb{N}\times\mathbb{Z}^{d-1}\bigr)\otimes\mathbb{C}^{N}$
where $d\in\mathbb{N}$ is the space dimension and $N\in\mathbb{N}$
is the number of internal degrees of freedom on each lattice site
(spin, or otherwise). Under the assumption of nearest-neighbor hopping
in the first axis a generic translation-invariant Hamiltonian on $l^{2}\bigl(\mathbb{Z}\bigr)\otimes L^{2}\bigl(\mathbb{T}^{d-1}\bigr)\otimes\mathbb{C}^{N}$
is a multiplication operator on $L^{2}\bigl(\mathbb{T}^{d-1}\bigr)\otimes\mathbb{C}^{N}$
given by: 
\begin{eqnarray}
\mathscr{H}\bigl(k^{\perp}\bigr) & = & \mathds{1}\otimes V\bigl(k^{\perp}\bigr)+S\otimes A\bigl(k^{\perp}\bigr)+S^{\ast}\otimes A\bigl(k^{\perp}\bigr)^{\ast}\label{eq:nearest neighbor translation invariant hamiltonian}
\end{eqnarray}
where $k^{\perp}$ ranges in $\mathbb{T}^{d-1}$ and $V:\mathbb{T}^{d-1}\to Herm_{N}\bigl(\mathbb{C}\bigr)$,
$A:\mathbb{T}^{d-1}\to Mat_{N\times N}\bigl(\mathbb{C}\bigr)$ the
matrices that define the Hamiltonian. $S$ is the right-shift operator
on $l^{2}\bigl(\mathbb{Z}\bigr)$: $\bigl(S\psi\bigr)\bigl(x\bigr)\equiv\psi\bigl(x-1\bigr)$. 
\begin{defn}
(\emph{Bulk Hamiltonians}) For the bulk system we may diagonalize
also the first axis to get a full multiplication operator on $L^{2}\bigl(\mathbb{T}^{d}\bigr)\otimes\mathbb{C}^{N}$:
\begin{eqnarray}
H\bigl(k\bigr) & = & V\bigl(k^{\perp}\bigr)+2\Re\bigl\{ A\bigl(k^{\perp}\bigr)\bigr\} \cos\bigl(k_{1}\bigr)+2\Im\bigl\{ A\bigl(k^{\perp}\bigr)\bigr\} \sin\bigl(k_{1}\bigr)\label{eq:nearest-neighbor translation invariant bulk hamiltonian}
\end{eqnarray}
where $k_{1}$ ranges in $\bigl[0,\,2\pi\bigr)\cong S^{1}$, $k=\bigl(k_{1},\,k^{\perp}\bigr)$.
Note $H$ has $N$ bands. Furthermore we require that $H$ has a spectral
gap, which means that we can two groups of bands which never intersect
for any value of $k$ (though within each group we allow intersections).
Stated as such it is clear that $H$ is equivalent to a map $P:\mathbb{T}^{d}\to Gr_{k}\bigl(\mathbb{C}^{N}\bigr)$
for some $k$ (which selects the gap) where $Gr_{k}\bigl(\mathbb{C}^{N}\bigr)$
is the Grassmannian manifold. In fact $P$ is the Fermi projection.
We require that $P$ is a continuous map which gives us a topology
on the space of gapped bulk Hamiltonians. This continuity is implied
by the locality of the underlying real-space Hamiltonian on $l^{2}\bigl(\mathbb{Z}^{d}\bigr)\otimes\mathbb{C}^{N}$.
\end{defn}

\begin{rem}
Note that for fixed $k^{\perp}$, this equation defines an abstract
ellipse in the space of self-adjoint $N\times N$ matrices, with the
plane of the ellipse defined by the two directions $2\Re\bigl\{ A\bigl(k^{\perp}\bigr)\bigr\} $
and $2\Im\bigl\{ A\bigl(k^{\perp}\bigr)\bigr\} $. This was the
great insight of \cite{Mong_Shivamoggi_PhysRevB.83.125109}.
\end{rem}

\begin{defn}
(\emph{Edge Hamiltonians}) The edge system with Dirichlet boundary
conditions is obtained from $\mathscr{H}$ after replacing the shift
operators $S$ by unilateral right shifts $S^{\sharp}$ on $l^{2}\bigl(\mathbb{N}\bigr)$.
We obtain an operator on $l^{2}\bigl(\mathbb{N}\bigr)\otimes L^{2}\bigl(\mathbb{T}^{d-1}\bigr)\otimes\mathbb{C}^{N}$
which is a multiplication operator on $L^{2}\bigl(\mathbb{T}^{d-1}\bigr)\otimes\mathbb{C}^{N}$
given by:
\begin{eqnarray*}
H^{\sharp}\bigl(k^{\perp}\bigr) & = & \mathds{1}\otimes V\bigl(k^{\perp}\bigr)+S^{\sharp}\otimes A\bigl(k^{\perp}\bigr)+S^{\sharp\ast}\otimes A\bigl(k^{\perp}\bigr)^{\ast}
\end{eqnarray*}
\end{defn}

\begin{rem}
As a rule, $H^{\sharp}$ has \emph{more} states than $H$, and in
particular we anticipate that generically it may not have a spectral
gap.
\end{rem}

\begin{defn}
(\emph{The Bulk-Edge Correspondence}) A topological invariant is a
continuous map $\mathcal{I}$ or $\mathcal{I}^{\sharp}$ from the
set of (bulk or edge) Hamiltonians into a discrete space, for us either
$\mathbb{Z}$ or $\mathbb{Z}_{2}$, and hence it is locally constant.
The bulk-edge correspondence is the fact that 
\begin{eqnarray*}
\mathcal{I}\bigl(H\bigr) & = & \mathcal{I}^{\sharp}\bigl(H^{\sharp}\bigr)
\end{eqnarray*}
for all bulk Hamiltonians $H$, where $H^{\sharp}$ is the edge Hamiltonian
induced by $H$ as above.
\end{defn}

\begin{rem}
The presentation so far has not referred to any symmetries, which
mean that we put further constraints on the Hamiltonians and define
different map $\mathcal{I}$ and $\mathcal{I}^{\sharp}$. This will
be done below as we consider concrete examples.
\end{rem}

\section{\label{sec:The-1D-Chiral}The 1D Chiral Case}

Following \cite{Schnyder_Ryu_Furusaki_Ludwig_PhysRevB.78.195125},
we define chiral systems of class $AIII$: We assume that $N\in2\mathbb{N}$,
so $N=2L$ for some $L\in\mathbb{N}$ and so we may write all matrices
in $2\times2$ blocks of $L\times L$ matrices. In this basis, we
define a chirality operator as 
\begin{eqnarray*}
\Pi & := & \begin{pmatrix}\mathds{1} & 0\\
0 & -\mathds{1}
\end{pmatrix}
\end{eqnarray*}
Note that $\Pi$ is local and acts only within the fibers of internal
degrees of freedom, hence there is no need to distinguish between
its variants before or after Bloch decomposition and we just use one
symbol for all.
\begin{defn}
A Hamiltonian $H$ is chiral-symmetric iff it anti-commutes with $\Pi$,
iff it has off-diagonal block form 
\begin{eqnarray*}
H & = & \begin{pmatrix}0 & T^{\ast}\\
T & 0
\end{pmatrix}
\end{eqnarray*}
for some $T$ which may not be necessarily self-adjoint.
\end{defn}

$ $
\begin{defn}
(\emph{Edge Chiral Topological Invariant}) We define $\mathcal{I}_{Chiral}^{\sharp}\bigl(H^{\sharp}\bigr)$
using $\dim\bigl(\ker\bigl(H^{\sharp}\bigr)\bigr)$ as follows:
It is the number of positive-chiral zero-energy states of $H^{\sharp}$
minus the negative-chiral zero-energy states.
\end{defn}

\begin{claim}
$\mathcal{I}_{Chiral}^{\sharp}\bigl(H^{\sharp}\bigr)$ is well-defined.
\begin{proof}
The following considerations are valid for both $H$ and $H^{\sharp}$
so we simply write $H$. If $H\psi=E\psi$, then $H\Pi\psi=-\Pi H\psi=-\Pi E\psi=-E\Pi\psi$.
So, $\psi$ is an eigenstate of energy $E$ iff $\Pi\psi$ is an eigenstate
of energy $-E$. These two are linearly independent if $E\neq0$.
Since $\Pi^{2}=\mathds{1}$, $\psi\pm\Pi\psi$ are two eigenstates
of $\Pi$ with eigenvalues $\pm1$. From this we also learn that $\ker\bigl(H\bigr)$
is an invariant subspace for $\Pi$, so that $\Pi$ may be diagonalized
within it. The conclusion is that outside of zero energy, states come
in pairs of opposite energy and chirality and that within the zero
energy eigenspace, chirality is a good quantum number. Hence by the
stability of the spectrum, the chirally-signed number of zero-energy
states of $H$ defines the topological invariant for the edge system,
which generally may have zero energy (in contrast to the bulk).
\end{proof}
\end{claim}

\begin{defn}
(\emph{Bulk Chiral Topological Invariant}) For the bulk, due to the
gap condition (the gap must be at zero for chirality) $H$ must be
invertible, so also $T$ must be. Hence we obtain a map $T:\mathbb{T}^{d}\to GL_{L}\bigl(\mathbb{C}\bigr)$
which has a well-defined winding number (after using the homotopy
equivalence $GL_{L}\bigl(\mathbb{C}\bigr)\sim\mathcal{U}\bigl(\mathbb{C}^{L}\bigr)$).
So $\mathcal{I}_{Chiral}\bigl(H\bigr):=Wind\bigl(T\bigr)$.
\end{defn}

The simplest non-trivial choice to make here is $L=1$ and $d=1$
which according to the periodic table gives systems classified by
a topological invariant in $\mathbb{Z}$. In this case the aforementioned
winding number is simply that of a map $T:S^{1}\to\mathbb{C}\backslash\Set{0}$.
If we further employ our nearest-neighbor assumption \eqref{eq:nearest-neighbor translation invariant bulk hamiltonian}
we get that $T$ traces an ellipse not intersecting zero in the complex
plane. Thus the nearest-neighbor approximation means there can only
be winding of $0$ or $\pm1$.
\begin{claim}
For $L=d=1$ nearest-neighbor models, we have $\mathcal{I}_{Chiral}\bigl(H\bigr)=\mathcal{I}_{Chiral}^{\sharp}\bigl(H^{\sharp}\bigr)$.
\begin{proof}
We may employ the results and notations of \cref{subsec:Dirac-Hamiltonians}
since $H$ is clearly a Dirac Hamiltonian (which is composed of the
Pauli matrices $\sigma_{1}$ and $\sigma_{2}$ only). In particular,
the result says that there is a zero energy edge state for $H^{\sharp}$
iff the ellipse traced by $h$ contains the origin.

Thus in the nearest neighbor approximation the indices may only be
in $\Set{0,\,\pm1}$. In the bulk this is clear: the degree of $h$,
where $h$ is an ellipse in $\mathbb{R}^{2}\backslash\Set{0}$, can
only be in $\Set{0,\,\pm1}$. In the edge this it is also clear that
for nearest-neighbor chiral models with only two levels there can
be at most kernel of dimension $1$, and by the results of \cref{subsec:Dirac-Hamiltonians},
if the ellipse doesn't include the origin, then its winding is zero
in which case $H^{\sharp}$ also has no kernel. So the goal is to
match the signs in the two possible cases. The claim is that iff the
ellipse $h$ goes counterclockwise then the edge mode that exists
has positive chirality and iff it goes clockwise then the edge mode
that exists has negative chirality.

This is seen by using the particular equations for the edge zero eigenvector
(again using the notations of \cref{subsec:Dirac-Hamiltonians}): 
\begin{eqnarray}
\bigl(\eta^{+}\bigl(\lambda\bigr)\sigma^{+}+\eta^{-}\bigl(\lambda\bigr)\sigma^{-}\bigr)u & = & 0\label{eq:zero energy edge eigenvector}
\end{eqnarray}
where $\sigma^{\pm}$ are defined in the usual way from $\sigma_{1}$
and $\sigma_{2}$: $\sigma^{\pm}:=\frac{1}{2}\bigl(\sigma_{1}\pm i\sigma_{2}\bigr)$
and the $\eta^{\pm}\bigl(\lambda\bigr)$ are now analytic continuation
of $h^{\pm}$ to complex $k$: 
\begin{eqnarray*}
\eta^{\pm}\bigl(\lambda\bigr) & \equiv & h^{\pm}\bigl(-i\log\bigl(\lambda\bigr)\bigr)\;\forall\lambda\in S^{1}
\end{eqnarray*}
From the analysis of \cref{subsec:Dirac-Hamiltonians}, we know that
the existence of a zero edge mode corresponds to either $\eta^{+}$
or $\eta^{-}$ having two zeros in the unit circle, but it's not possible
that each has only one. As a result \eqref{eq:zero energy edge eigenvector}
becomes 
\[
\begin{cases}
\eta^{s}\bigl(\lambda_{1}\bigr)\sigma^{s}u & =0\\
\eta^{s}\bigl(\lambda_{2}\bigr)\sigma^{s}u & =0
\end{cases}
\]
with $s\in\Set{\pm}$. In the first case $u=\begin{pmatrix}u^{\bigl(1\bigr)}\\
0
\end{pmatrix}$ for some $u^{\bigl(1\bigr)}\in\mathbb{C}\backslash\Set{0}$ and
in the second case $u=\begin{pmatrix}0\\
u^{\bigl(2\bigr)}
\end{pmatrix}$ for some $u^{\bigl(2\bigr)}\in\mathbb{C}\backslash\Set{0}$. But
these are exactly eigenvectors of $\sigma_{3}=\Pi$, our chirality
operator, with eigenvalues $+1$ and $-1$ respectively. Now recall
in \cref{subsec:Dirac-Hamiltonians} we found that iff $h$ goes counter-clockwise
then it is $\eta^{+}$ which has two zeros in the unit circle whereas
iff $h$ goes clockwise then it is $\eta^{-}$.
\end{proof}
\end{claim}

\begin{rem}
It should be stated that for such systems (with more relaxed assumptions)
\cite{PSB_2016} had already given a proof of the bulk-edge correspondence,
and in \cite{Graf_Shapiro_Bulk-Edge_2017} we investigate the case
of strong disorder for such systems, which is more general than the
proof of \cite{PSB_2016}.
\end{rem}

\section{\label{sec:The-Integer-Quantum-Hall}The IQHE Case}

The IQHE ($d=2$) belongs to class $A$ in \cite{Schnyder_Ryu_Furusaki_Ludwig_PhysRevB.78.195125}
which means it has no symmetry constraints. The simplest gapped system
thus has $N=2$. Such models are automatically Dirac (as in \eqref{eq:Dirac_Bulk_Hamiltonian})
with the Pauli matrices.
\begin{defn}
(\emph{Bulk IQHE Topological Invariant}) A bulk Hamiltonian is thus
equivalent to a map $P:\mathbb{T}^{2}\to Gr_{1}\bigl(\mathbb{C}^{2}\bigr)\cong S^{2}$
which has a well-defined degree (equal to the first Chern number $Ch_{1}$
of the $\mathbb{T}^{2}$-vector bundle $E$ it induces). We define
$\mathcal{I}_{QH}\bigl(H\bigr):=\deg\bigl(P\bigr)$.
\end{defn}

$ $
\begin{defn}
(\emph{Edge IQHE Topological Invariant}) We define $\mathcal{I}_{QH}^{\sharp}\bigl(H^{\sharp}\bigr)$
as the signed number of crossings of the edge states which cross the
Fermi energy $E_{F}\in\mathbb{R}$ in the (single) bulk gap.
\end{defn}

\begin{center}
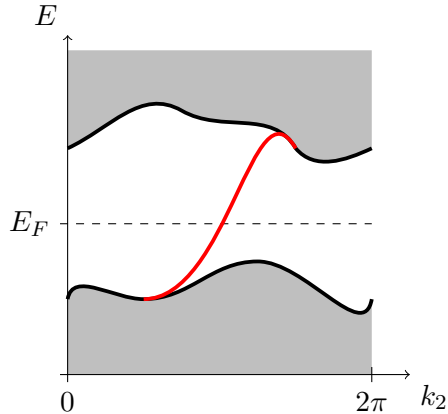
\begin{figure}[h]
    \begin{center}\begin{tikzpicture}[scale=1]     \path [fill=lightgray] (0,1) to [out=80,in=180] (1,1) to [out=0,in=180] (2.5,1.5) to [out=0,in=-100] (4,1) to (4,0) to (0,0);     \draw [line width=0.05cm] (0,1) to [out=80,in=180] (1,1) to [out=0,in=180] (2.5,1.5) to [out=0,in=-100] (4,1);     \path [fill=lightgray] (0,3) to [out=25,in=150] (1.5,3.5) to [out=-30,in=120] (3,3) to [out=-50,in=-155] (4,3) to (4,4.3) to (0,4.3);     \draw [line width=0.05cm] (0,3) to [out=25,in=150] (1.5,3.5) to [out=-30,in=120] (3,3) to [out=-50,in=-155] (4,3);                   \draw [thin, ->] (-0.1,0) -- (4.5,0);     \draw [thin, ->] (0, -0.1) -- (0, 4.5);     \node [below right] at (4.5,0) {$k_2$};     \node [above left] at (0, 4.5) {$E$};     \draw [thin] (4, -0.1) -- (4, 0.1);     \node [below] at (4, -0.1) {$2\pi$};     \node [below] at (0, -0.1) {$0$};          \draw [dashed] (-0.1, 2) -- (4,2);     \node [left] at (-0.1, 2) {$E_F$};     \draw [line width=0.05cm, red] (1, 1) to [out=0, in=120] (3, 3);          \end{tikzpicture}\end{center}

\caption{The bulk spectrum is in the shaded area, and the discrete edge spectrum
is in red. In this configuration, $\mathcal{I}_{QH}^{\sharp}\bigl(H^{\sharp}\bigr)=-1$
because there is one crossing point with positive slope.}
\end{figure}
\par\end{center}
\begin{rem}
\label{rem:edge index if fermi energy has non zero slope}We may choose
to take an arbitrary fiducial line $E_{F}\bigl(k_{2}\bigr)$ instead
of a constant $E_{F}$, such that $E_{F}'\bigl(k_{2}\bigr)\neq0$. 
\end{rem}

\subsection{Proof of the Bulk-Edge Correspondence for Singular Hopping Matrices}

The proof presented in this section does not rely on \cref{subsec:Dirac-Hamiltonians},
as it makes the
\begin{assumption}
\label{assu:Singular Hopping Matrix}The hopping matrix $A$ in \cref{eq:nearest-neighbor translation invariant bulk hamiltonian}
is of the form 
\[
A=\begin{pmatrix}a_{11} & 0\\
a_{21} & 0
\end{pmatrix}
\]
\end{assumption}

Under this assumption, we have $A\psi_{0}=\begin{pmatrix}a_{11}\bigl(\psi_{0}\bigr)_{1}\\
a_{21}\bigl(\psi_{0}\bigr)_{1}
\end{pmatrix}$, so it suffices to require merely $\bigl(\psi_{0}\bigr)_{1}\stackrel{!}{=}0$
as the boundary condition of the edge eigenvalue problem, and $\bigl(\psi_{0}\bigr)_{2}$
can in fact stay unconstrained. This possibility allows us to avoid
having to find linear combinations of generalized Bloch solutions
as in \cref{subsec:Edge-Spectrum-via-Complex-Momentum} and we can
look for non-zero edge solutions simply by imposing the boundary condition
\begin{equation}
\psi_{0}\stackrel{!}{=}\begin{pmatrix}0\\
\ast
\end{pmatrix}\label{eq:edge boundary condition for two-band model singular hopping matrix}
\end{equation}
We still don't know the edge spectrum, but we do know that at the
points of incipience, it is equal to the bulk spectrum.
\begin{example}
One possible example for when \cref{assu:Singular Hopping Matrix}
holds is if we take a model that has no internal degrees of freedom,
but has periodicity of \emph{two} lattice sites in the $1$-direction.
This is the case for example of the Harper equation that \cite{Hatsugai_PhysRevLett.71.3697}
first analyzed.
\end{example}

\begin{claim}
\label{claim:bulk index signed number of preimages of the north pole}The
following is a simple fact about the degree of the map stated without
proof: $\mathcal{I}_{QH}\bigl(H\bigr)$ is the signed number of points
from $\mathbb{T}^{2}$ that reach the north pole of $S^{2}$ via the
map $\hat{h}:\mathbb{T}^{2}\to S^{2}$ given by $\hat{h}:=\frac{h}{\norm{h}}$
where $h$ is as in \eqref{eq:Dirac_Bulk_Hamiltonian}.
\end{claim}

\begin{defn}
Define the supremum of the lower energy band as 
\[
E_{l,sup}\bigl(k_{2}\bigr):=\sup\bigl(\Set{E_{lower}\bigl(k\bigr)|k_{1}\in S^{1}}\bigr)
\]
\end{defn}

\begin{defn}
Denote the \emph{discrete }edge spectrum as $E^{\sharp}\bigl(k_{2}\bigr)$
for all $k_{2}\in S^{1}$. We know there is only one energy eigenvalue
in the gap for the edge state, from \cref{claim:sign_of_edge_spectrum_in_dirac_Hamiltonians}.
\end{defn}

\begin{claim}
\label{claim:edge quantum hall index for two band models singular hopping matrix}$\mathcal{I}_{QH}^{\sharp}\bigl(H^{\sharp}\bigr)$
is given by the signed number of degeneracy points between $E^{\sharp}\bigl(k_{2}\bigr)$
and $E_{l,sup}^{B}\bigl(k_{2}\bigr)$, where the sign is obtained
via the relative slope of the $E^{\sharp}\bigl(k_{2}\bigr)$ and
$E_{l,sup}\bigl(k_{2}\bigr)$.
\begin{proof}
We may take without loss of generality the Fermi energy to be infinitesimally
close to $E_{l,sup}\bigl(k_{2}\bigr)$: 
\[
E_{F}\bigl(k_{2}\bigr):=E_{l,sup}\bigl(k_{2}\bigr)
\]
and then \cref{rem:edge index if fermi energy has non zero slope}
gives exactly the above definition.
\end{proof}
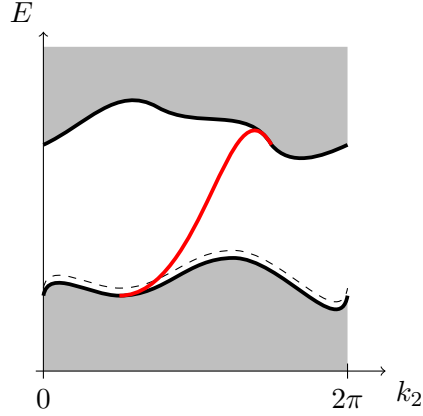
\begin{figure}[h]
\begin{center}\begin{tikzpicture}      
\path [fill=lightgray] (0,1) to [out=80,in=180] (1,1) to [out=0,in=180] (2.5,1.5) to [out=0,in=-100] (4,1) to (4,0) to (0,0);      
\draw [line width=0.05cm] (0,1) to [out=80,in=180] (1,1) to [out=0,in=180] (2.5,1.5) to [out=0,in=-100] (4,1);      
\path [fill=lightgray] (0,3) to [out=25,in=150] (1.5,3.5) to [out=-30,in=120] (3,3) to [out=-50,in=-155] (4,3) to (4,4.3) to (0,4.3);     
\draw [line width=0.05cm] (0,3) to [out=25,in=150] (1.5,3.5) to [out=-30,in=120] (3,3) to [out=-50,in=-155] (4,3);                   
\draw [thin, ->] (-0.1,0) -- (4.5,0);      
\draw [thin, ->] (0, -0.1) -- (0, 4.5);      
\node [below right] at (4.5,0) {$k_2$};      
\node [above left] at (0, 4.5) {$E$};      
\draw [thin] (4, -0.1) -- (4, 0.1);      
\node [below] at (4, -0.1) {$2\pi$};      
\node [below] at (0, -0.1) {$0$};           
  
\draw [dashed] (0,1.1) to [out=80,in=180] (1,1.1) to [out=0,in=180] (2.5,1.6) to [out=0,in=-100] (4,1.1);     \draw [line width=0.05cm, red] (1, 1) to [out=0, in=120] (3, 3);           
\end{tikzpicture}\end{center}

\caption{Another possibility to find the edge index is to set the fiducial
line (in this picture the dashed line) infinitesimally close to $E_{l,sup}^{B}\bigl(k_{2}\bigr)$
(in this picture the thick lower line).}
\end{figure}

\end{claim}

\begin{claim}
For $N\in S^{2}$ the north pole, $\hat{h}^{-1}\bigl(\Set{N}\bigr)$
is the set of points in $\mathbb{T}^{2}$ such that $E_{l,sup}\bigl(k_{2}\bigr)$
is degenerate with $E^{\sharp}\bigl(k_{2}\bigr)$.
\begin{proof}
In order to be able to compare the two descriptions, we first compute
the eigenvectors of the bulk system \emph{before} we allow for the
possibility that $k_{1}$ is complex-valued (this generalization follows
from \cref{subsec:Edge-Spectrum-via-Complex-Momentum}). Consequently,
$h\in\mathbb{R}^{3}$ for the purpose of this computation.

Let $k\in\mathbb{T}^{2}$ be given. The eigensystem equation is given
by 
\[
\begin{pmatrix}+h_{3} & h_{1}-ih_{2}\\
h_{1}+ih_{2} & -h_{3}
\end{pmatrix}\begin{pmatrix}v_{1}\,^{\bigl(n\bigr)}\\
v_{2}\,^{\bigl(n\bigr)}
\end{pmatrix}=\bigl(-1\bigr)^{n}\norm{h}\begin{pmatrix}v_{1}\,^{\bigl(n\bigr)}\\
v_{2}\,^{\bigl(n\bigr)}
\end{pmatrix}
\]
which gives an eigenvector $\begin{pmatrix}v_{1}\,^{\bigl(n\bigr)}\\
v_{2}\,^{\bigl(n\bigr)}
\end{pmatrix}$ corresponding to the eigenvalue $E_{n}=\bigl(-1\bigr)^{n}\norm{h}$
for $\begin{cases}
n=1 & lower\\
n=2 & upper
\end{cases}$.

From this equation two equations follow for $v_{1}$ and $v_{2}$:
\[
\begin{cases}
\bigl(h_{3}-\bigl(-1\bigr)^{n}\norm{h}\bigr)v_{1}\,^{\bigl(n\bigr)}+\bigl(h_{1}-ih_{2}\bigr)v_{2}\,^{\bigl(n\bigr)} & =0\\
\bigl(h_{1}+ih_{2}\bigr)v_{1}\,^{\bigl(n\bigr)}+\bigl(-h_{3}-\bigl(-1\bigr)^{n}\norm{h}\bigr)v_{2}\,^{\bigl(n\bigr)} & =0
\end{cases}
\]
dividing through $\norm{h}$ (which is never zero by hypothesis) we
get:
\[
\begin{cases}
\bigl(\hat{h}_{3}-\bigl(-1\bigr)^{n}\bigr)v_{1}\,^{\bigl(n\bigr)}+\bigl(\hat{h}_{1}-i\hat{h}_{2}\bigr)v_{2}\,^{\bigl(n\bigr)} & =0\\
\bigl(\hat{h}_{1}+i\hat{h}_{2}\bigr)v_{1}\,^{\bigl(n\bigr)}+\bigl(-\hat{h}_{3}-\bigl(-1\bigr)^{n}\bigr)v_{2}\,^{\bigl(n\bigr)} & =0
\end{cases}
\]

Note that, as before, even though we started with a general point
$\begin{pmatrix}h_{0}\\
h
\end{pmatrix}\in\mathbb{R}^{4}$ such that $\norm{h}\neq0$, what matters for the eigenvectors is
only the associated point $\hat{h}\in S^{2}$.
\begin{enumerate}
\item Case 1: $\hat{h}_{3}=1$ (the north pole, where $\hat{h}_{1}=\hat{h}_{2}=0$)
\end{enumerate}
\begin{itemize}
\item Then we obtain 
\[
\begin{cases}
\bigl(1-\bigl(-1\bigr)^{n}\bigr)v_{1}\,^{\bigl(n\bigr)} & =0\\
\bigl(-1-\bigl(-1\bigr)^{n}\bigr)v_{2}\,^{\bigl(n\bigr)} & =0
\end{cases}
\]
\item Then for $n=1$, $v_{2}\,^{\bigl(1\bigr)}$ is free and $v_{1}\,^{\bigl(1\bigr)}$
must be zero, so that we obtain that the general eigenvector corresponding
to $E_{1}$ is given by $\begin{pmatrix}0\\
\alpha
\end{pmatrix}$ for some $\alpha\in\mathbb{C}\backslash\bigl\{ 0\bigr\} $.
\item For $n=2$, $v_{1}\,^{\bigl(2\bigr)}$ is free and $v_{2}\,^{\bigl(2\bigr)}$
must be zero, so that the general eigenvector corresponding to $E_{2}^{B}$
is given by $\begin{pmatrix}\alpha\\
0
\end{pmatrix}$ for some $\alpha\in\mathbb{C}\backslash\bigl\{ 0\bigr\} $.
\end{itemize}
\begin{enumerate}
\item Case 2: $\hat{h}_{3}=-1$ (the south pole, where $\hat{h}_{1}=\hat{h}_{2}=0$)
\end{enumerate}
\begin{itemize}
\item Then we obtain 
\[
\begin{cases}
\bigl(-1-\bigl(-1\bigr)^{n}\bigr)v_{1}\,^{\bigl(n\bigr)} & =0\\
\bigl(+1-\bigl(-1\bigr)^{n}\bigr)v_{2}\,^{\bigl(n\bigr)} & =0
\end{cases}
\]
\item For $n=1$, $v_{1}\,^{\bigl(1\bigr)}$ is free and $v_{2}\,^{\bigl(1\bigr)}$
must be zero, so that we obtain that the general eigenvector corresponding
to $E_{1}$ is given by $\begin{pmatrix}\alpha\\
0
\end{pmatrix}$ for some $\alpha\in\mathbb{C}\backslash\bigl\{ 0\bigr\} $.
\item For $n=2$, $v_{2}\,^{\bigl(2\bigr)}$ is free and $v_{1}\,^{\bigl(2\bigr)}$
must be zero, so that the general eigenvector corresponding to $E_{2}$
is given by $\begin{pmatrix}0\\
\alpha
\end{pmatrix}$ for some $\alpha\in\mathbb{C}\backslash\bigl\{ 0\bigr\} $.
\end{itemize}
\begin{enumerate}
\item Case 3: $\hat{h}_{3}\notin\Set{\pm1}$ (where either $\hat{h}_{1}\neq0$
or $\hat{h}_{2}\neq0$)
\end{enumerate}
\begin{itemize}
\item Then we obtain 
\[
\begin{cases}
v_{1}\,^{\bigl(n\bigr)} & =\frac{-\hat{h}_{1}+i\hat{h}_{2}}{\hat{h}_{3}-\bigl(-1\bigr)^{n}}v_{2}\,^{\bigl(n\bigr)}\\
v_{1}\,^{\bigl(n\bigr)} & =\frac{\hat{h}_{3}+\bigl(-1\bigr)^{n}}{\hat{h}_{1}+i\hat{h}_{2}}v_{2}\,^{\bigl(n\bigr)}
\end{cases}
\]
\item The two equations are the same up to multiplication by a nonzero constant
complex number and so the general eigenvector associated with $E_{n}$
is given by $\begin{pmatrix}\frac{-\hat{h}_{1}+i\hat{h}_{2}}{\hat{h}_{3}-\bigl(-1\bigr)^{n}}\alpha\\
\alpha
\end{pmatrix}$ for some $\alpha\in\mathbb{C}\backslash\bigl\{ 0\bigr\} $. Observe
that the two components of this vector will never be zero, because
we assume $\hat{h}_{3}\notin\Set{\pm1}$.
\end{itemize}
The final conclusion from this analysis is that the non-normalized
(but normalizable) eigenvectors for the bulk system are given, up
to multiplication by non-zero complex factors by, 
\[
\begin{cases}
\begin{cases}
\psi_{lower}= & \begin{pmatrix}0\\
1
\end{pmatrix}\\
\psi_{upper}= & \begin{pmatrix}1\\
0
\end{pmatrix}
\end{cases} & \hat{h}_{3}=1\\
\begin{cases}
\psi_{lower}= & \begin{pmatrix}1\\
0
\end{pmatrix}\\
\psi_{upper}= & \begin{pmatrix}0\\
1
\end{pmatrix}
\end{cases} & \hat{h}_{3}=-1\\
\begin{cases}
\psi_{lower}= & \begin{pmatrix}\frac{-\hat{h}_{1}+i\hat{h}_{2}}{\hat{h}_{3}+1}\\
1
\end{pmatrix}\\
\psi_{upper}= & \begin{pmatrix}\frac{-\hat{h}_{1}+i\hat{h}_{2}}{\hat{h}_{3}-1}\\
1
\end{pmatrix}
\end{cases} & \hat{h}_{3}\notin\Set{\pm1}
\end{cases}
\]
We can already see that the edge boundary condition \cref{eq:edge boundary condition for two-band model singular hopping matrix}
on the bulk eigenvectors are only fulfilled on special points on the
sphere: on the north pole only for $\psi_{lower}$ whereas on the
south pole only for $\psi_{upper}$. 

Hence we may conclude that for these points on the sphere, we exactly
have degeneracy between edge energy eigenvalues and bulk energy eigenvalues,
because these solutions \emph{are} solutions of the bulk Hamiltonian
$H$ (with real values of $k$) yet they also obey the boundary conditions
of the edge, and as we saw in \cref{subsec:Edge-Spectrum-via-Complex-Momentum},
solutions of $H$ which obey the edge boundary conditions are solutions
of $H^{\sharp}$.
\end{proof}
\end{claim}

\begin{rem}
We were able to find these degeneracy points using the bulk Hamiltonian
and the edge boundary conditions alone, with no analysis of the edge
system nor its actual discrete spectrum. In fact, had the word ``signed''
not been used in the definition of $\mathcal{I}_{QH}$ (with or without
$\sharp$), the correspondence proof would have been done at this
point. Hence the missing fact from the correspondence proof is the
matching of the signs, so that the counting would indeed be the same.
All effort done from this point onward will be invested to that end. 
\end{rem}

Let $k^{D}\in h{}^{-1}\bigl(\Set{N}\bigr)$ be such a degeneracy
point between $E_{l,sup}\bigl(k_{2}\bigr)$ and $E^{\sharp}\bigl(k_{2}\bigr)$.
We will show that both signs of the edge and the bulk agree for $k^{D}$
and thus complete the proof that 
\begin{equation}
\mathcal{I}_{QH}\bigl(H\bigr)=\mathcal{I}_{QH}^{\sharp}\bigl(H^{\sharp}\bigr)\label{eq:two band model singular hopping matrices bulk edge correspondence}
\end{equation}
\begin{claim}
The equation $h_{1}=ih_{2}$ determines the complex value of $k_{1}$
of the edge solution in terms of $k_{2}$.
\begin{proof}
In general, to find $E^{\sharp}\bigl(k_{2}\bigr)$, we can solve
the eigensystem of $H$, but assuming that $k_{1}$ can take on complex
values in the upper plane (so that the edge wavefunctions decay exponentially
into the bulk, as we expect from edge states at the zeroth site),
and impose the edge boundary $\psi_{0}\stackrel{!}{=}0$ and that
$E^{\sharp}\bigl(k_{2}\bigr)\in\mathbb{R}$. However, those boundary
conditions cannot be imposed on the same wavefunctions (that is, eigenvectors)
of $H\bigl(k_{2}\bigr)$ where we assumed $k_{1}\in\mathbb{R}$.
We need to ``re-solve'' the eigensystem allowing for $\Im\bigl(k_{1}\bigr)>0$
and only then impose the boundary conditions. As a result, now we
allow $h\in\mathbb{C}^{3}$. As such the matrix $\sum_{j=1}^{3}h_{j}\sigma_{j}$
is no longer Hermitian and our eigenvalues are not \emph{necessarily}
real: 
\[
E_{1,2}^{\sharp}=\pm\sqrt{h_{1}+h_{2}+h_{3}}
\]
The eigenvectors are given by the two equations which come from the
eigenvalue equation $H\bigl(k\bigr)\begin{pmatrix}v_{1}\,^{\bigl(n\bigr)}\\
v_{2}\,^{\bigl(n\bigr)}
\end{pmatrix}=\bigl(+\bigl(-1\bigr)^{n}\sqrt{h_{1}+h_{2}+h_{3}}\bigr)\begin{pmatrix}v_{1}\,^{\bigl(n\bigr)}\\
v_{2}\,^{\bigl(n\bigr)}
\end{pmatrix}$: 
\[
\begin{cases}
\bigl(h_{3}-\bigl(-1\bigr)^{n}\sqrt{h_{1}+h_{2}+h_{3}}\bigr)v_{1}\,^{\bigl(n\bigr)}+\bigl(h_{1}-ih_{2}\bigr)v_{2}\,^{\bigl(n\bigr)} & =0\\
\bigl(h_{1}+ih_{2}\bigr)v_{1}\,^{\bigl(n\bigr)}+\bigl(-h_{3}-\bigl(-1\bigr)^{n}\sqrt{h_{1}+h_{2}+h_{3}}\bigr)v_{2}\,^{\bigl(n\bigr)} & =0
\end{cases}
\]
We don't actually need to compute the eigenvectors, but rather, only
check when they obey the boundary conditions, that is, when it would
follow from the equations that $v_{1}\,^{\bigl(n\bigr)}=0$ and $v_{2}\,^{\bigl(n\bigr)}\neq0$,
following \cref{eq:edge boundary condition for two-band model singular hopping matrix}.
To that end, we get the equations: 
\[
\begin{cases}
\bigl(h_{3}-\bigl(-1\bigr)^{n}\sqrt{h_{1}+h_{2}+h_{3}}\bigr) & \neq0\\
\bigl(h_{1}-ih_{2}\bigr) & =0\\
\bigl(h_{1}+ih_{2}\bigr) & \neq0\\
\bigl(-h_{3}-\bigl(-1\bigr)^{n}\sqrt{h_{1}+h_{2}+h_{3}}\bigr) & =0
\end{cases}
\]
These conditions are fulfilled when $h_{1}=ih_{2}$: $\bigl(ih_{2}\bigr)^{2}+\bigl(h_{2}\bigr)^{2}=0$,
and when $E^{e}=-h_{3}$. Thus, the equation $h_{1}=ih_{2}$ determines
the complex value of $k_{1}$ in the edge in terms of $k_{2}$. 
\end{proof}
\end{claim}

If we had explicit expressions for $h_{1}$ and $h_{2}$ we could
already look at the expression $E^{\sharp}\bigl(k_{2}\bigr)$ and
compute its slope in the vicinity of $k^{D}$. Because we don't, we
will make an approximation at $k^{D}+\delta$ with $\delta_{2}\in\mathbb{R}$,
$\delta_{1}\in\mathbb{C}$ (in upper plane for decaying solution)
and $\bigl|\delta_{i}\bigr|\ll1\forall i\in J_{2}$. 
\begin{claim}
The edge spectrum near $E_{l,sup}\bigl(k_{D}\bigr)$ is obtained
by taking $\Re\bigl\{ \delta_{1}\bigr\} =0$.
\begin{proof}
To get the edge spectrum near $k^{D}$, we plug into $E_{lower}^{B}$
complex values of $k_{1}$ such that the result is real, and that
the corresponding eigenstates obey the boundary conditions and decay.
Observe that $k_{1}^{D}$ is an extremal point of $E_{lower}^{B}\bigl(k\bigr)$
for fixed $k_{2}$, and as such, $\partial_{1}E_{lower}^{B}\bigl(k^{D}\bigr)=0$.
Thus 
\[
E_{lower}^{B}\bigl(k_{1}^{D}+\delta_{1},\,k_{2}^{D}+\delta_{2}\bigr)\approx E_{lower}^{B}\bigl(k^{D}\bigr)+\partial_{2}E_{lower}^{B}\bigl(k^{D}\bigr)\delta_{2}+\bigl[\partial_{1}^{2}E_{lower}^{B}\bigl(k^{D}\bigr)\bigr]\delta_{1}^{2}
\]
As a result we see that the only way for $E_{lower}^{B}\bigl(k_{1}^{D}+\delta_{1},\,k_{2}^{D}+\delta_{2}\bigr)$
to be real (and thus, to represent the edge energy) is to have $\Re\bigl(\delta_{1}\bigr)=0$
(and so $\delta_{1}$ is \emph{purely} imaginary in the upper complex
plane).
\end{proof}
\end{claim}

\begin{claim}
The signs of the edge index count and the bulk index count exactly
match, proving \cref{eq:two band model singular hopping matrices bulk edge correspondence}.
\begin{proof}
Now that we have all the ingredients, we may proceed as follows:
\begin{itemize}
\item The condition that $h_{1}=ih_{2}$ translates to (in the vicinity
of $k^{D}$): 
\[
\underbrace{h_{1}\bigl(k^{D}\bigr)}_{0}+i\bigl[\partial_{1}h_{1}\bigl(k^{D}\bigr)\bigr]\Im\bigl\{ \delta_{1}\bigr\} +\bigl[\partial_{2}h_{1}\bigl(k^{D}\bigr)\bigr]\delta_{2}\stackrel{!}{\approx}i\bigl\{ \underbrace{h_{2}\bigl(k^{D}\bigr)}_{0}+i\bigl[\partial_{1}h_{2}\bigl(k^{D}\bigr)\bigr]\Im\bigl\{ \delta_{1}\bigr\} +\bigl[\partial_{2}h_{2}\bigl(k^{D}\bigr)\bigr]\delta_{2}\bigr\} 
\]
From which we obtain that 
\[
\begin{cases}
\bigl[\partial_{1}h_{1}\bigl(k^{D}\bigr)\bigr]\Im\bigl\{ \delta_{1}\bigr\}  & =\bigl[\partial_{2}h_{2}\bigl(k^{D}\bigr)\bigr]\delta_{2}\\
\bigl[\partial_{2}h_{1}\bigl(k^{D}\bigr)\bigr]\delta_{2} & =-\bigl[\partial_{1}h_{2}\bigl(k^{D}\bigr)\bigr]\Im\bigl\{ \delta_{1}\bigr\} 
\end{cases}
\]
or in shorthand notation 
\[
\begin{cases}
h_{1,1}\Im\bigl\{ \delta_{1}\bigr\}  & =h_{2,2}\delta_{2}\\
h_{2,1}\Im\bigl\{ \delta_{1}\bigr\}  & =-h_{1,2}\delta_{2}
\end{cases}
\]
These equation imply 
\[
\begin{cases}
h_{1,1}\,^{2}\Im\bigl\{ \delta_{1}\bigr\}  & =h_{1,1}h_{2,2}\delta_{2}\\
h_{2,1}\,^{2}\Im\bigl\{ \delta_{1}\bigr\}  & =-h_{2,1}h_{1,2}\delta_{2}
\end{cases}
\]
which in turn implies 
\[
\Im\bigl\{ \delta_{1}\bigr\} =\frac{h_{1,1}h_{2,2}-h_{1,2}h_{2,1}}{h_{1,1}\,^{2}+h_{2,1}\,^{2}}\delta_{2}
\]
\item We know that $\Im\bigl\{ \delta_{1}\bigr\} >0$ (that's the condition
for a decaying solution). Following \cref{claim:edge quantum hall index for two band models singular hopping matrix},
$sgn\bigl(\delta_{2}\bigr)$ gives us a way to determine the sign
of the degeneracy point for the count of $\mathcal{I}_{QH}^{\sharp}\bigl(H^{\sharp}\bigr)$.
Indeed, iff $\delta_{2}>0$ we get a plus sign for the count of the
edge index because $E^{\sharp}\bigl(k_{D}\bigr)$ grows rightwards,
which is happens iff the relative slope of $E^{\sharp}\bigl(\bigl(k_{D}\bigr)_{2}\bigr)$
with $E_{l,sup}\bigl(\bigl(k_{D}\bigr)_{2}\bigr)$ is positive,
because $E^{\sharp}\bigl(\bigl(k_{D}\bigr)_{2}+\delta_{2}\bigr)$
must be above $E_{l,sup}\bigl(\bigl(k_{D}\bigr)_{2}+\delta_{2}\bigr)$
by definition; this is in agreement with \cref{rem:edge index if fermi energy has non zero slope}.
The analog argument holds if $\delta_{2}<0$. Observe $\delta_{2}=0$
is not possible by the stability of the edge spectrum.
\item Use the fact that $\partial_{i}\hat{h}_{j}=\frac{1}{\norm{h}}\bigl[\partial_{i}h_{j}-\bigl\langle \hat{h},\,\partial_{i}h\bigr\rangle \hat{h}_{j}\bigr]$
to conclude that $\partial_{i}\hat{h}_{j}\bigl(k^{D}\bigr)=\frac{1}{h_{3}\bigl(k^{D}\bigr)}\partial_{i}h_{j}\forall j\in J_{2}$
where $h_{3}\bigl(k^{D}\bigr)>0$. Thus, the signs are preserved
and we can safely replace the condition 
\[
\partial_{1}\hat{h}_{1}\bigl(k^{D}\bigr)\partial_{2}\hat{h}_{2}\bigl(k^{D}\bigr)-\partial_{1}\hat{h}_{2}\bigl(k^{D}\bigr)\partial_{2}\hat{h}_{1}\bigl(k^{D}\bigr)>0
\]
by 
\[
\partial_{1}h_{1}\bigl(k^{D}\bigr)\partial_{2}h_{2}\bigl(k^{D}\bigr)-\partial_{1}h_{2}\bigl(k^{D}\bigr)\partial_{2}h_{1}\bigl(k^{D}\bigr)>0
\]
which is what we had in \cref{claim:bulk index signed number of preimages of the north pole}.
\item Thus we see that the two signs \emph{exactly match }because when $\delta_{2}>0$,
$\hat{h}\bigl(k_{D}\bigr)$ is orientation preserving \emph{and}
the count of the edge index is $+1$, and the corresponding statement
for $\delta_{2}<0$.
\end{itemize}
\end{proof}
\end{claim}

\begin{center}
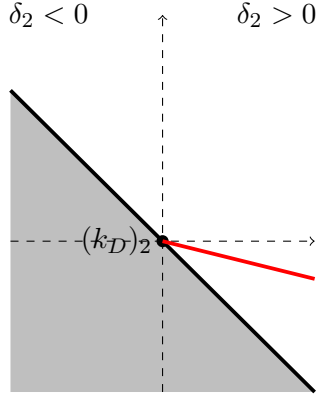
\begin{figure}[t]
\begin{center}\begin{tikzpicture}   \node at (-1.5, 3) {$\delta_2<0$}; \node at (1.5, 3) {$\delta_2>0$};
\path [fill=lightgray] (-2,2) to (2,-2) to (-2,-2) to (-2,2); \draw [line width=0.05cm] (-2,2) -- (2,-2); \fill (0,0) circle [radius=0.08cm]; \draw [dashed, ->] (0,-2) -- (0,3); \draw [dashed, ->] (-2,0) -- (2,0); \draw [red, line width=0.05cm] (0,0) -- (2,-0.5);      \node [below, left] at (0,0) {$(k_D)_2$};  \end{tikzpicture}\end{center}

\caption{A linear zoom near $\bigl(k_{D}\bigr)_{2}$. Even though $E^{\sharp}\bigl(\bigl(k_{D}\bigr)_{2}\bigr)$
(red line) has negative slope, relative to $E_{l,sup}\bigl(\bigl(k_{D}\bigr)_{2}\bigr)$
(thick black line) it has positive slope. This has to be the case
when $\delta_{2}>0$ because the the red line \emph{must} always be
above the thick black line.}
\end{figure}
\par\end{center}

\subsection{\label{subsec:Bulk-Edge-Correspondence-for-General-2-band-Hamiltonian}Bulk-Edge
Correspondence for a General Two-Band Hamiltonian}

We now relax \cref{assu:Singular Hopping Matrix} so that $A$ from
\eqref{eq:nearest-neighbor translation invariant bulk hamiltonian}
could be any $2\times2$ matrix. We present our version of the proof
in \cite{Mong_Shivamoggi_PhysRevB.83.125109} to compare to the one
above in the same setting. Most of the work is already contained in
\cref{claim:Dirac_Hamiltonians_Edge_Spectrum}.
\begin{claim}
\label{claim:chern_number_2_band_is_number_of_great_circles}In a
nearest-neighbor approximation, if for a fixed $k_{2}$, $\bigl.\hat{h}\bigl(k\bigr)\bigr|_{k_{2}}$
is a loop on $S^{2}$ parametrized by $k_{1}$, then $Ch_{1}\bigl(E\bigr)$
is given by the signed number of times $\bigl.\hat{h}\bigl(k\bigr)\bigr|_{k_{2}}$
crosses as a great circle on $S^{2}$, where the sign is (with the
notation in \cref{claim:Dirac_Hamiltonians_Edge_Spectrum}):
\begin{itemize}
\item Positive if $\bigl(\hat{e}^{r}\times\hat{e}^{i}\bigr)\cdot\hat{e}^{\perp}$
went from $-1$ before crossing to $+1$ after crossing.
\item Negative if $\bigl(\hat{e}^{r}\times\hat{e}^{i}\bigr)\cdot\hat{e}^{\perp}$
went from $+1$ before crossing to $-1$ after crossing.
\end{itemize}
\begin{proof}
Even though the curve $\bigl.h\bigl(k\bigr)\bigr|_{k_{2}}$ is planar,
$\bigl.\hat{h}\bigl(k\bigr)\bigr|_{k_{2}}$ is not necessarily a
circle when projected on $S^{2}$, and so the area $\bigl.\hat{h}\bigl(k\bigr)\bigr|_{k_{2}}$
encloses is not strictly speaking a cap on $S^{2}$, so we'll refer
to it as a ``cap''. We write the formula in \cite{ASS_83_Homotopy_and_Quantization}
as:
\begin{eqnarray*}
Ch_{1}\bigl(E\bigr) & = & \frac{1}{4\pi}\int_{0}^{2\pi}f\bigl(k_{2}\bigr)\dif{k_{2}}
\end{eqnarray*}
with $f\bigl(k_{2}\bigr):=\int_{0}^{2\pi}\hat{h}\bigl(k\bigr)\cdot\bigl\{ \bigl[\partial_{k_{1}}\hat{h}\bigl(k\bigr)\bigr]\times\bigl[\partial_{k_{2}}\hat{h}\bigl(k\bigr)\bigr]\bigr\} \dif{k_{1}}$.
The integral in $Ch_{1}\bigl(E\bigr)$ gives the total surface covered
on the sphere covered by $\mathbb{T}^{2}$, so that the function $f$
is the rate of change of area-covering, as we change $k_{2}$. That
is, $\int_{k_{2}=k_{2}^{\bigl(1\bigr)}}^{k_{2}=k_{2}^{\bigl(2\bigr)}}f\bigl(k_{2}\bigr)dk_{2}$
gives the area on the sphere enclosed between the two curves $\bigl.\hat{h}\bigl(k\bigr)\bigr|_{k_{2}^{\bigl(1\bigr)}}$
and $\bigl.\hat{h}\bigl(k\bigr)\bigr|_{k_{2}^{\bigl(2\bigr)}}$.
We may thus define the anti-derivative of $f\bigl(k_{2}\bigr)$ as
$F\bigl(k_{2}\bigr)$ and so $Ch_{1}\bigl(E\bigr)=\frac{1}{4\pi}\bigl[F\bigl(2\pi\bigr)-F\bigl(0\bigr)\bigr]$.
$F$ is such that $F'\bigl(k_{2}\bigr)=f\bigl(k_{2}\bigr)$, but
we have the freedom to define $F\bigl(k_{2}\bigr)$ with an arbitrary
constant. Define the constant as follows: Let $k_{2}^{\bigl(0\bigr)}\in S^{1}$
be given such that $\hat{e}^{\perp}\bigl(k_{2}^{\bigl(0\bigr)}\bigr)\neq0$.
Define $F\bigl(k_{2}^{\bigl(0\bigr)}\bigr)$ as the area of the
``cap'' enclosed by the loop $\bigl.\hat{h}\bigl(k\bigr)\bigr|_{k_{2}^{\bigl(0\bigr)}}$,
where the ``cap'' is the one the vector $-\bigl(\hat{e}^{r}\times\hat{e}^{i}\bigr)$
points towards. If such a $k_{2}^{\bigl(0\bigr)}\in S^{1}$ does
not exist then the claim is automatically satisfied as there are no
crossings at all, and $Ch_{1}\bigl(E\bigr)=0$ indeed because if
$\bigl.\hat{h}\bigl(k\bigr)\bigr|_{k_{2}}$ covers any area, in
order to obey the boundary conditions it will cover a negative area
of the same amount, totalling in zero. $F\bigl(k_{2}\bigr)$ thus
is the amount of area swept on $S^{2}$ going from $k_{2}^{\bigl(0\bigr)}$
to some $k_{2}$. As a side note, using the Guass-Bonnet theorem we
could verify this directly: 
\[
Area\bigl(k_{2}\bigr)=2\pi\underbrace{\chi}_{1}-\int_{\bigl.\hat{h}\bigl(k\bigr)\bigr|_{k_{2}}}k_{g}\bigl(s\bigr)ds
\]
where $k_{g}\bigl(s\bigr)$ is the curvature along $\bigl.\hat{h}\bigl(k\bigr)\bigr|_{k_{2}}$,
$s$ is the arc-length parametrization and $\chi$ is the Euler characteristic
of a closed disc. 

At any rate, now $Ch_{1}\bigl(E\bigr)$ can be computed as the \emph{signed}
number of times $F$ crosses the lines $2\pi\mathbb{Z}$, because
$\bigl.\hat{h}\bigl(k\bigr)\bigr|_{k_{2}^{\bigl(0\bigr)}}=\bigl.\hat{h}\bigl(k\bigr)\bigr|_{k_{2}^{\bigl(0\bigr)}+2\pi}$
as $k_{2}\in S^{1}$, so that $F\bigl(k_{2}^{\bigl(0\bigr)}+2\pi\bigr)=n\cdot F\bigl(k_{2}^{\bigl(0\bigr)}\bigr)$
for some $n\in\mathbb{Z}$: 
\[
Ch_{1}\bigl(E\bigr)=\frac{1}{4\pi}\bigl[F\bigl(k_{2}^{\bigl(0\bigr)}+2\pi\bigr)-F\bigl(k_{2}^{\bigl(0\bigr)}\bigr)\bigr]=n
\]
In order to have $n\neq0$, the loop $\bigl.\hat{h}\bigl(k\bigr)\bigr|_{k_{2}}$
must cross as a great circle, and when this happens, $F$ is a multiple
of $2\pi$, as that is the area of exactly half of $S^{2}$. Furthermore,
at the crossings, $\hat{e}^{\perp}=0$ because then the ellipse is
not offset perpendicularly from the origin. So there, $\bigl(\hat{e}^{r}\times\hat{e}^{i}\bigr)\cdot\hat{e}^{\perp}=0$. 

The choice of ``cap'' we made for $F\bigl(k_{2}^{\bigl(0\bigr)}\bigr)$
ensures that the sign of the crossing must be counted as $+1$ if
$\bigl(\hat{e}^{r}\times\hat{e}^{i}\bigr)\cdot\hat{e}^{\perp}$ went
from being negative to positive and $-1$ if $\bigl(\hat{e}^{r}\times\hat{e}^{i}\bigr)\cdot\hat{e}^{\perp}$
went from being positive to negative:
\begin{itemize}
\item If it happens that $\bigl(\hat{e}^{r}\bigl(k_{2}^{\bigl(0\bigr)}\bigr)\times\hat{e}^{i}\bigl(k_{2}^{\bigl(0\bigr)}\bigr)\bigr)\cdot\hat{e}^{\perp}\bigl(k_{2}^{\bigl(0\bigr)}\bigr)>0$,
then $-\bigl(\hat{e}^{r}\bigl(k_{2}^{\bigl(0\bigr)}\bigr)\times\hat{e}^{i}\bigl(k_{2}^{\bigl(0\bigr)}\bigr)\bigr)$
points towards the larger ``cap'' defined by $\bigl.\hat{h}\bigl(k\bigr)\bigr|_{k_{2}^{\bigl(0\bigr)}}$.
\item If it happens that $\bigl(\hat{e}^{r}\bigl(k_{2}^{\bigl(0\bigr)}\bigr)\times\hat{e}^{i}\bigl(k_{2}^{\bigl(0\bigr)}\bigr)\bigr)\cdot\hat{e}^{\perp}\bigl(k_{2}^{\bigl(0\bigr)}\bigr)<0$,
then $-\bigl(\hat{e}^{r}\bigl(k_{2}^{\bigl(0\bigr)}\bigr)\times\hat{e}^{i}\bigl(k_{2}^{\bigl(0\bigr)}\bigr)\bigr)$
points towards the smaller ``cap'' defined by $\bigl.\hat{h}\bigl(k\bigr)\bigr|_{k_{2}^{\bigl(0\bigr)}}$.
\end{itemize}
In either case we have that as $F$ crosses the lines $2\pi\mathbb{Z}$
with positive slope, we necessarily have that $\bigl(\hat{e}^{r}\bigl(k_{2}\bigr)\times\hat{e}^{i}\bigl(k_{2}\bigr)\bigr)\cdot\hat{e}^{\perp}\bigl(k_{2}\bigr)$
goes from being $-1$ to $+1$. As $F$ crosses the lines $2\pi\mathbb{Z}$
with negative slope, $\bigl(\hat{e}^{r}\bigl(k_{2}\bigr)\times\hat{e}^{i}\bigl(k_{2}\bigr)\bigr)\cdot\hat{e}^{\perp}\bigl(k_{2}\bigr)$
goes from $+1$ to $-1$.
\end{proof}

\end{claim}

\begin{cor}
For nearest-neighbour two-band models we have $\mathcal{I}_{QH}\bigl(H\bigr)=\mathcal{I}_{QH}^{\sharp}\bigl(H^{\sharp}\bigr)$.
\begin{proof}
Without loss of generality, we set $E_{F}=0$. Then, using \cref{claim:Dirac_Hamiltonians_Edge_Spectrum}
we know that there is a zero-energy edge state iff the ellipse $\bigl.h\bigl(k\bigr)\bigr|_{k_{2}}$
contains the origin, and at these points $k_{2}\in S^{1}$, $\bigl.\hat{h}\bigl(k\bigr)\bigr|_{k_{2}}$
is a great circle on $S^{2}$.

Furthermore, from \cref{claim:sign_of_edge_spectrum_in_dirac_Hamiltonians}
we know that the sign of the edge energy around zero is given by $\bigl(\hat{e}^{r}\times\hat{e}^{i}\bigr)\cdot\hat{e}^{\perp}$,
so that following the definition \cref{eq:Edge_Hall_Conductance}:
\begin{itemize}
\item $-\frac{\bigl(E^{\sharp}\bigr)'\bigl(k_{2}\bigr)}{\bigl|\bigl(E^{\sharp}\bigr)'\bigl(k_{2}\bigr)\bigr|}=1$
if $\bigl(\hat{e}^{r}\times\hat{e}^{i}\bigr)\cdot\hat{e}^{\perp}$
went from being $-1$ before the crossing to $+1$ after the crossing.
\item $-\frac{\bigl(E^{\sharp}\bigr)'\bigl(k_{2}\bigr)}{\bigl|\bigl(E^{\sharp}\bigr)'\bigl(k_{2}\bigr)\bigr|}=-1$
if $\bigl(\hat{e}^{r}\times\hat{e}^{i}\bigr)\cdot\hat{e}^{\perp}$
went from being $+1$ before the crossing to $-1$ after the crossing.
\end{itemize}
\end{proof}
\end{cor}

\section{\label{sec:The-2D-Time-Reversal}The 2D Time-Reversal Invariant Case}

We consider the case when $d=2$. Time-reversal $\Theta$ is an anti-unitary
map $\Theta^{2}=-\mathds{1}$ which takes a state above $k$ to a
vector above $-k$. The Hamiltonian $H$ is called time-reversal invariant
(TRI from now on) iff 
\begin{eqnarray}
H\bigl(-k\bigr) & = & \Theta H\bigl(k\bigr)\Theta^{-1}\label{eq:TRI for Bloch Decomposed}
\end{eqnarray}
Then according to \cite{Kitaev2009} the index should be $\mathbb{Z}_{2}$-valued. 

Kramers theorem says then that at those special points where $k=-k$
(henceforth this subset of $\mathbb{T}^{d}$ is called $TRIM$ for
time-reversal invariant momenta), the spectrum of $H\bigl(k\bigr)$
is at least two-fold degenerate. This implies that here in order to
get the simplest non-trivial gapped case we must choose $N=4$, which
we do. Note that unlike the previous IQHE analysis, for $N=4$ a generic
gapped self-adjoint Hamiltonian is actually not necessarily of the
form \eqref{eq:Dirac_Bulk_Hamiltonian}\textendash there are in general
sixteen gamma matrices and only five of which may anti-commute simultaneously.
\begin{defn}
(\emph{Edge Kane-Mele Topological Invariant}) We define:
\begin{equation}
\mathcal{I}_{KM}^{\sharp}\bigl(H^{\sharp}\bigr):=\frac{1}{2}\mathcal{I}_{QH}^{\sharp}\bigl(H^{\sharp}\bigr)\mod2\label{eq:Edge_Kane_Mele_Index}
\end{equation}
Note this makes sense using the above properties of $H^{\sharp}$
which ensure $\mathcal{I}_{QH}^{\sharp}\bigl(H^{\sharp}\bigr)\in2\mathbb{Z}$.
\end{defn}

Despite the first introduction of the Kane-Mele index being in \cite{Kane_Mele_2005_PhysRevLett.95.146802},
we follow instead the equivalent definition of \cite{PhysRevB.74.195312}
(equation 3.25).\footnote{Note that \cite{PhysRevB.74.195312} provides a proof for the equivalence
of the definition we use with the definition of \cite{Kane_Mele_2005_PhysRevLett.95.146802}.}

Define a matrix 
\begin{equation}
w_{ij}\bigl(k\bigr):=\bigl\langle \psi_{i}\bigl(-k\bigr),\,\Theta\psi_{j}\bigl(k\bigr)\bigr\rangle \label{eq:w matrix of kane-mele index}
\end{equation}
where $\Set{\psi_{i}\bigl(k\bigr)}_{i}$ is a set of eigenstates
of $H\bigl(k\bigr)$ corresponding to the occupied states, each of
which is chosen \emph{continuously} throughout $\mathbb{T}^{2}$.
Thus, $\Set{\psi_{j}}_{j\in Occupied}$ is a continuous section in
the occupied frame bundle over $\mathbb{T}^{2}$.
\begin{claim}
Such a global smooth choice of $\Set{\psi_{i}\bigl(k\bigr)}_{i}$
is always possible for time-reversal-invariant systems, due to the
fact that TRI forces $\mathcal{I}_{QH}\bigl(H\bigr)=0$.
\end{claim}

\begin{claim}
$w\bigl(k\bigr)=-\bigl[w\bigl(-k\bigr)\bigr]^{T}$
\end{claim}

\begin{rem}
As a result, we see that $Pf\bigl[w\bigl(k\bigr)\bigr]$ is defined
$\forall k\in TRIM$, as at such points $w\bigl(k\bigr)$ is anti-symmetric
and, by hypothesis, there is always an even number of occupied bands.
\end{rem}

\begin{defn}
\label{def:The Kane-Mele Bulk Index}\emph{(Bulk Kane-Mele Topological
Invariant}) Define 
\begin{equation}
\mathcal{I}_{KM}\bigl(H\bigr):=\frac{1}{i\pi}\log\bigl(\prod_{k\in TRIM}\frac{\sqrt{\det\bigl[w\bigl(k\bigr)\bigr]}}{Pf\bigl[w\bigl(k\bigr)\bigr]}\bigr)\label{eq:Bulk-Z2_Invariant_Definition}
\end{equation}
\end{defn}

\begin{rem}
Naively, it would seem that $\mathcal{I}_{KM}\bigl(H\bigr)$ is always
zero, due to $\det\bigl[A\bigr]=\bigl(Pf\bigl[A\bigr]\bigr)^{2}$.
However, care must be taken with the branch of $\sqrt{\cdot}$ that
is chosen, which has to be done globally on $\mathbb{T}^{2}$. As
a result, even though the formula does not \emph{explicitly} require
one to compute $\det\bigl[w\bigl(k\bigr)\bigr]$ outside of $TRIM\subset\mathbb{T}^{2}$,
in order to make a continuous choice of $\sqrt{\det\bigl[w\bigl(k\bigr)\bigr]}$,
knowledge of $\det\bigl[w\bigl(k\bigr)\bigr]$ along paths in $\mathbb{T}^{2}\backslash TRIM$
connecting points in $TRIM$ is necessary. It is in this part that
the assumption of $\Set{\psi_{j}}_{j}$ being a smooth section will
be used. \\
It should also be noted that $\bigl(\prod_{k\in TRIM}\frac{\sqrt{\det\bigl[w\bigl(k\bigr)\bigr]}}{Pf\bigl[w\bigl(k\bigr)\bigr]}\bigr)\in\Set{\pm1}$
and so $\mathcal{I}_{KM}\bigl(H\bigr)\in\mathbb{\mathbb{Z}}_{2}$
indeed (usually there is a confusion between $\Set{0,\,1}$ and $\Set{\pm1}$).
\end{rem}

The most general Bloch decomposed Hamiltonian may be written as 
\[
H=\sum_{\bigl(i,\,j\bigr)\in\bigl(\mathbb{Z}_{4}\bigr)^{2}}d_{i,\,j}\Gamma_{i,\,j}
\]
with $\Gamma_{i,\,j}\equiv\sigma_{i}\otimes\sigma_{j}$ the 16 gamma
matrices. This is \emph{not} the form of \eqref{eq:Dirac_Bulk_Hamiltonian}. 

We may without loss of generality assume that $d_{0,0}\bigl(k\bigr)=0$,
using the fact that it would merely shift the bulk spectrum by a constant
amount at each point, but cannot close the gap. Note that we do allow
for the two lower bands intersect, and similarly for the two upper
bands. The following is stated without proof:
\begin{claim}
\label{claim:TREI_Coefficients_are_even}If we choose $\Theta=\Gamma_{0,\,2}\mathcal{C}$
where $\mathcal{C}$ is complex-conjugation and mapping between the
fibers of $k$ to $-k$, then with $TREI:=\Set{\bigl(2,\,1\bigr),\,\bigl(2,\,2\bigr),\,\bigl(2,\,3\bigr),\,\bigl(0,\,0\bigr),\,\bigl(1,\,0\bigr),\,\bigl(3,\,0\bigr)}$,
it follows due to \cref{eq:TRI for Bloch Decomposed} that 
\[
d_{i,\,j}\bigl(-k\bigr)=d_{i,\,j}\bigl(k\bigr)\,\,\,\forall\bigl(i,\,j\bigr)\in TREI
\]
and 
\[
d_{i,\,j}\bigl(-k\bigr)=-d_{i,\,j}\bigl(k\bigr)\,\,\,\forall\bigl(i,\,j\bigr)\notin TREI
\]

\end{claim}

\begin{claim}
\label{claim:Special Formula For Kane-Mele index for Dirac Hamiltonians}Let
$A\subset\bigl(\mathbb{Z}_{4}\bigr)^{2}\backslash\Set{\bigl(0,\,0\bigr)}$
be given such that $\bigl\{ \Gamma_{i,\,}\Gamma_{j}\bigr\} =2\delta_{i,j}\forall\bigl(i,\,j\bigr)\in A^{2}$.
If $d_{j}\bigl(k\bigr)=0\quad\forall j\in A^{c}$ then \cref{def:The Kane-Mele Bulk Index}
reduces to 
\begin{equation}
\mathcal{I}_{KM}\bigl(H\bigr)=\begin{cases}
0 & \bigl|A\cap TREI\bigr|>1\\
\frac{1}{i\pi}\log\bigl(\prod_{k\in TRIM}sgn\bigl(d_{e}\bigl(k\bigr)\bigr)\bigr) & A\cap TREI=\Set{e}
\end{cases}\label{eq:New Formula for Kane-Mele Index}
\end{equation}
\begin{proof}
We divide into two cases according to \cref{eq:New Formula for Kane-Mele Index}:
\begin{enumerate}
\item Case 1: $A\cap TREI=\Set{e}$.

The following proof is a generalization of one given in \cite{Fu_Kane_2007_PhysRevB.76.045302}
for the case of spacetime-inversion symmetric systems, where here
$\Gamma_{e}$ takes the role of space-inversion. Since it is merely
a generalization, we only include a sketch of the proof here.
\begin{claim*}
$\bigl[H\bigl(k\bigr),\,\Gamma_{e}\Theta\bigr]=0$.\label{claim:even index tr is a symmetry of hamiltonian}
\end{claim*}
$ $
\begin{claim*}
$H\bigl(-k\bigr)=\Gamma_{e}H\bigl(k\bigr)\bigl(\Gamma_{e}\bigr)^{-1}$
\end{claim*}
\begin{defn*}
Define a $2\times2$ matrix by its components 
\[
v_{m,\,n}\bigl(k\bigr)=\bigl\langle \psi_{m}\bigl(k\bigr),\,\Gamma_{e}\Theta\psi_{n}\bigl(k\bigr)\bigr\rangle \quad\forall\bigl(m,n\bigr)\in\bigl(J_{2}\bigr)^{2}
\]
where $\psi_{n}\bigl(k\bigr)\in\mathbb{C}^{4}$ is the $n$th eigenstate
of $H\bigl(k\bigr)$ and we are assuming that the $1$ and $2$ bands
are the occupied ones.
\end{defn*}
\begin{claim*}
$v\bigl(k\bigr)$ is an anti-symmetric matrix.
\end{claim*}
\begin{cor*}
The Pfaffian of $v\bigl(k\bigr)$ is defined and it is given by 
\[
Pf\bigl[v\bigl(k\bigr)\bigr]=v_{12}\bigl(k\bigr)
\]
\end{cor*}
\begin{claim*}
$\bigl|Pf\bigl[v\bigl(k\bigr)\bigr]\bigr|=1$.\label{claim:abs val of Pfaffian is one}
\begin{proof}
Using \cref{claim:even index tr is a symmetry of hamiltonian} we have
that $\psi_{m}\bigl(k\bigr)$ and $\Gamma_{e}\Theta\psi_{m}\bigl(k\bigr)$
are both eigenstates of the same energy.
\begin{claim*}
$\Gamma_{e}\Theta\psi_{m}\bigl(k\bigr)$ is linearly independent
of $\psi_{m}\bigl(k\bigr)$.
\end{claim*}
As a result, because there are only two states, we conclude that $\psi_{1}\bigl(k\bigr)=e^{i\theta\bigl(k\bigr)}\Gamma_{e}\Theta\psi_{2}\bigl(k\bigr)$
for some $\theta:\mathbb{T}^{2}\to\mathbb{R}$.

\end{proof}
\begin{claim*}
\label{claim:WLOG pfaffian is one}Without loss of generality, we
may assume that we are in such a gauge such that $Pf\bigl[v\bigl(k\bigr)\bigr]=1$.
\begin{proof}
Under the action of a gauge transformation of the form 
\[
\psi_{n}\bigl(k\bigr)\stackrel{G}{\mapsto}e^{i\alpha\bigl(k\bigr)\delta_{n,1}}\psi_{n}\bigl(k\bigr)
\]
$Pf\bigl[v\bigl(k\bigr)\bigr]$ transforms as 
\begin{eqnarray*}
Pf\bigl[v\bigl(k\bigr)\bigr] & \stackrel{G}{\mapsto} & e^{-i\alpha\bigl(k\bigr)}Pf\bigl[v\bigl(k\bigr)\bigr]
\end{eqnarray*}
Thus, according to \cref{claim:abs val of Pfaffian is one}, $Pf\bigl[v\bigl(k\bigr)\bigr]=e^{-i\theta\bigl(k\bigr)}$
and so if we pick $\alpha=-\theta$ then we can make sure that $Pf\bigl[v\bigl(k\bigr)\bigr]=1$
for all $k\in\mathbb{T}^{2}$.
\end{proof}
\end{claim*}
$ $
\begin{claim*}
Using \cref{claim:WLOG pfaffian is one} we then have that 
\[
\det\bigl[w\bigl(k\bigr)\bigr]=1
\]
where $w$ was defined in \cref{eq:w matrix of kane-mele index}.
\end{claim*}
Note that because $H\bigl(k\bigr)$ is comprised of anti-commuting
gamma matrices, we may employ \cref{eq:Dirac_Bulk_Hamiltonian} to
obtain that the two lower bands are always-degenerate with energy
\[
E_{lower}\bigl(k\bigr)=-\norm{d\bigl(k\bigr)}
\]

Next, due to the fact that $\det\bigl[w\bigl(k\bigr)\bigr]=1$ for
all $k\in\mathbb{T}^{2}$, we don't need to worry about picking the
right branch of $\sqrt{\det\bigl[w\bigl(k\bigr)\bigr]}$ continuously
over $\mathbb{T}^{2}$ and thus \cref{eq:Bulk-Z2_Invariant_Definition}
reduces to 
\[
\mathcal{I}_{KM}\bigl(H\bigr):=\frac{1}{i\pi}\log\bigl(\prod_{k\in TRIM}\frac{1}{Pf\bigl[w\bigl(k\bigr)\bigr]}\bigr)
\]
and our only concern is to compute $Pf\bigl[w\bigl(k\bigr)\bigr]$
at $k\in TRIM$. But At $k\in TRIM$, $H\bigl(k\bigr)=d_{e}\bigl(k\bigr)\Gamma_{e}$
because of $d_{i}\bigl(-k\bigr)=-d_{i}\bigl(k\bigr)\forall i\notin TREI$.
Thus we have 
\begin{eqnarray*}
d_{e}\bigl(k\bigr)\Gamma_{e}\psi_{n}\bigl(k\bigr) & = & -\bigl|d_{e}\bigl(k\bigr)\bigr|\psi_{n}\bigl(k\bigr)\\
\Gamma_{e}\psi_{n}\bigl(k\bigr) & = & -sgn\bigl(d_{e}\bigl(k\bigr)\bigr)\psi_{n}\bigl(k\bigr)
\end{eqnarray*}
Then we calculate $w\bigl(k\bigr)$ at $k\in TRIM$:$w_{mn}\bigl(k\bigr)=-sgn\bigl(d_{e}\bigl(k\bigr)\bigr)v_{mn}\bigl(k\bigr)$
so that $Pf\bigl[w\bigl(k\bigr)\bigr]=w_{12}\bigl(k\bigr)=-sgn\bigl(d_{e}\bigl(k\bigr)\bigr)$,
and the result follows using the fact that $\bigl|TRIM\bigr|\in2\mathbb{N}$.

Note that in particular, $d_{e}\bigl(k\bigr)\neq0\forall k\in TRIM$
because otherwise $d=0$ at such a point, and then the gap closes,
which by hypothesis is not possible.
\end{claim*}
\item Case 2: $\bigl|A\cap TREI\bigr|>1$.
\begin{enumerate}
\item Case 2.1: $A\subseteq TREI$.

In this case we have that all $d_{j}\bigl(k\bigr)$ coefficients
are symmetric, and as a result, the TRI condition $H\bigl(-k\bigr)=\Theta H\bigl(k\bigr)\Theta^{-1}$
becomes $\bigl[H\bigl(k\bigr),\,\Theta\bigr]=0$ by virtue of $H\bigl(-k\bigr)=H\bigl(k\bigr)$.
\begin{defn*}
Define a matrix $u$ by its components 
\[
u_{mn}\bigl(k\bigr):=\bigl\langle \psi_{m}\bigl(k\bigr),\,\Theta\psi_{n}\bigl(k\bigr)\bigr\rangle \forall\bigl(m,\,n\bigr)\in\bigl(J_{2}\bigr)^{2}
\]
\end{defn*}
It is easy to verify just as above that $u$ is anti-symmetric (unlike
$w$, but like $v$) on the whole of $\mathbb{T}^{2}$.
\begin{claim*}
$\bigl|Pf\bigl[u\bigl(k\bigr)\bigr]\bigr|=1$.
\end{claim*}
Again, we pick a gauge in which $Pf\bigl[u\bigl(k\bigr)\bigr]=1$
and similarly we have again that $u\bigl(-k\bigr)=w\bigl(k\bigr)\overline{u\bigl(k\bigr)}\bigl[w\bigl(k\bigr)\bigr]^{T}$
so that $\det\bigl[w\bigl(k\bigr)\bigr]=1$ here as well. On $k\in TRIM$,
$w=u$ and so $Pf\bigl[w\bigl(k\bigr)\bigr]=1$ for all $k\in TRIM$.
As a result, in this case we find that 
\[
\mathcal{I}_{KM}\bigl(H\bigr)=0
\]
\item Case 2.2: $H\bigl(k\bigr)=H_{0}\forall k\in\mathbb{T}^{2}$.

Here we actually still have $H\bigl(k\bigr)=H\bigl(-k\bigr)$ trivially
and so the preceding case covers this one.
\item Case 2.3: None of the above.
\begin{claim}
If $\bigl|A\cap TREI\bigr|>1$ then $H\bigl(k\bigr)$ is nullhomotopic
(to be justified in \cref{rem:Geometric Justification for Special Kane-Mele Formula}).
\end{claim}

As a result, we may adiabatically transform $H\bigl(k\bigr)$ to
a constant and so $\mathcal{I}_{KM}\bigl(H\bigr)=0$ again.
\end{enumerate}
\end{enumerate}
\end{proof}
\end{claim}

\begin{figure}[t]
\begin{center}\begin{tikzpicture}             
\path [fill=lightgray] (-3.14,0) to (3.14,0) to (3.14,3.14) to (-3.14,3.14) to (-3.14,0); \draw [thin, ->] (-3.3,0) -- (3.3, 0); \draw [thin, ->](0,-3.3) -- (0,3.3); \draw [thick, red] (-3.14,0) -- (3.14,0); \draw [thick, blue] (-3.14,3.14) -- (3.14,3.14); \draw [line width=0.1cm, cyan, dashed] (-3.14,0) -- (-3.14,3.14); \draw [line width=0.1cm, cyan, dashed] (3.14,0) -- (3.14,3.14); \fill (0,0) circle [radius=0.06cm]; \fill (3.14,0) circle [radius=0.06cm]; \fill (0,3.14) circle [radius=0.06cm]; \fill (3.14,3.14) circle [radius=0.06cm]; \fill (-3.14,3.14) circle [radius=0.06cm]; \fill (-3.14,0) circle [radius=0.06cm]; \fill (-3.14,-3.14) circle [radius=0.06cm]; \fill (0,-3.14) circle [radius=0.06cm]; \fill (3.14,-3.14) circle [radius=0.06cm]; \draw [thick, red] (1,-0.1) to (1,0.1);  \draw [thick, red] (-1,-0.1) to (-1,0.1); \node [below] at (1,-0.1) {$k$}; \node [below] at (-1,-0.1) {$-k$}; \fill [gray] (2,1.5) circle [radius=0.06cm]; \fill [gray] (-2,-1.5) circle [radius=0.06cm]; \node [above] at (2,1.5) {$\tilde{k}$}; \node [below] at (-2,-1.5) {$-\tilde{k}$};
\end{tikzpicture}\end{center}

\caption{Instead of specifying Hamiltonians on the whole of $\mathbb{T}^{2}$,
we only need to specify Hamiltonians in the shaded area. Within that
area, TRI does not need to be enforced. The dashed cyan lines are
stiched together as before to form a cylinder, and the red and blue
circles must still obey TRI, because their $k$ and $-k$ partners
are both included in the effective brillouin zone. }
\end{figure}
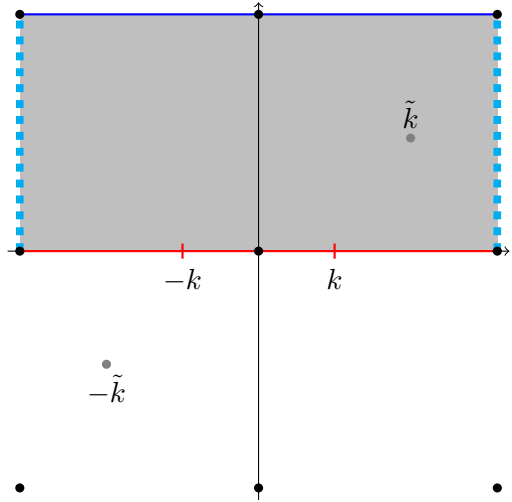

\begin{figure}[h]
\begin{center}\begin{tikzpicture}[scale=0.5]   \coordinate (ll) at (-3,-2);    \coordinate (lr) at (3,-2);    \coordinate (ul) at (-3,2);    \coordinate (ur) at (3,2);    \draw [thin, ->] (-3,-2) -- (-3, 3);    \node [left] at (-3,3) {$k_2$};
   \shade [shading angle=90] (ll) arc (-180:-60:3cm and .75cm) -- +(0,4) arc (-60:-180:3cm and .75cm) -- cycle;    \shade [shading angle=270] (lr) arc (0:-60:3cm and .75cm) -- +(0,4) arc (-60:0:3cm and .75cm) -- cycle;    \draw [thick] (ll) arc (-180:0:3cm and .75cm) -- (ur) arc (0:-180:3cm and .75cm) -- cycle;    \fill [blue, thick, shading angle=30] (ul) arc (-180:180:3cm and .75cm);       \draw [thick,red] (ll) arc (-180:0:3cm and .75cm) -- (lr);       \draw [thick,red,dashed] (ll) arc (180:0:3cm and .75cm) -- (lr);          \draw [thick,blue] (ul) arc (-180:0:3cm and .75cm) -- (ur);       \draw [thick,blue] (ul) arc (180:0:3cm and .75cm) -- (ur);          \node [left] at (ll) {$(0,0)$};       \node [right] at (lr) {$(\pi,0)$};    \node [left] at (ul) {$(0,\pi)$};       \node [right] at (ur) {$(\pi,\pi)$};
\end{tikzpicture}\end{center}

\caption{The resulting cylinder from the effective Brillouin zone. The upper
and lower discs are not included, but their boundary circles (red
and blue) are, and only on them do we enforce TRI.}
\end{figure}
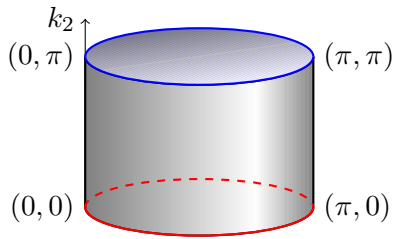
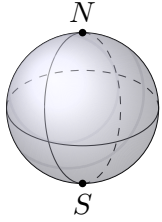
\begin{figure}[h]
\begin{center}\begin{tikzpicture}     \draw (-1,0) arc (180:360:1cm and 0.5cm);     \draw[dashed] (-1,0) arc (180:0:1cm and 0.5cm);     \draw (0,1) arc (90:270:0.5cm and 1cm);     \draw[dashed] (0,1) arc (90:-90:0.5cm and 1cm);     \draw (0,0) circle (1cm);     \shade[ball color=blue!10!white,opacity=0.20] (0,0) circle (1cm);     \fill (0,1) circle [radius=0.05cm];      \fill (0,-1) circle [radius=0.05cm];      \node [above] at (0,1) {$N$};      \node [below] at (0,-1) {$S$}; \end{tikzpicture}\end{center}

\caption{A path on $S^{4}$ (here depicted as $S^{2}$ for simplicity) which
has to start at $N$ and end at $S$ cannot be deformed into one that
starts and ends at $N$.}
\end{figure}
\begin{rem}
\label{rem:Geometric Justification for Special Kane-Mele Formula}The
formula in \cref{claim:Special Formula For Kane-Mele index for Dirac Hamiltonians}
can also be justified using a geometric argument.
\begin{proof}
First note that without loss of generality we may assume that $\bigl|E\bigl(k\bigr)\bigr|=1$,
as such a smooth change to $H$ would not close the gap. As a result,
$\norm{d\bigl(k\bigr)}=1$ at all points $k\in\mathbb{T}^{2}$. Also
note that due to the Clifford algebra, the maximal value of $\bigl|A\bigr|$
is $5$ so that in the general case we therefore have $d\bigl(k\bigr)\in S^{4}$
and this defines a map $\mathbb{T}^{2}\ni k\mapsto d\bigl(k\bigr)\in S^{4}$
which we seek to classify, up to homotopies which preserve TRI. Note
that in this setting, preserving TRI means preserving the evenness
or oddness of the componenets of $d$.

Our way to make this classification follows the beginning of \cite{Moore_Balents_2007_PhysRevB.75.121306}
closely: we work with the EBZ (``effective brillouin zone''), where
due to \eqref{eq:TRI for Bloch Decomposed}, $H\bigl(k\bigr)$ can
be fully specified on merely \emph{half} of $\mathbb{T}^{2}$, with
special boundary conditions. In the interior of the half-torus there
is no need to employ \eqref{eq:TRI for Bloch Decomposed} because
there is no partner $-k$ included in the EBZ for each given $k\in EBZ$.
That is, except at the boundary circles. 

Concretely, we pick the $EBZ$ half torus to work on as $k_{1}\in\bigl[-\pi,\,\pi\bigr]$
and $k_{2}\in\bigl[0,\,\pi\bigr]$. Then, the upper-right quadrant
$k\in\bigl[0,\,\pi\bigr]\times\bigl[0,\,\pi\bigr]$ specifies what
happens on the lower-left quadrant $k\in\bigl[-\pi,\,0\bigr]\times\bigl[-\pi,\,0\bigr]$
and similarly the upper-left quadrant specifies what happens on the
lower-right quadrant, all thanks to \eqref{eq:TRI for Bloch Decomposed}.
We still need to stich together the two boundaries at the lines $k_{1}=\pi$
and $k_{1}=-\pi$ so that we are left with a cylinder: $\bigl(k_{1},\,k_{2}\bigr)\in S^{1}\times\bigl[0,\,\pi\bigr]$
where $k_{1}$ can be thought of as the angle parameter of the cylinder
and $k_{2}$ is the ``height'' parameter of the cylinder. Notice
that the two circles at the top and the bottom of $S^{1}\times\bigl[0,\,\pi\bigr]$
\emph{are} included in the EBZ.

We now turn to the boundary conditions: Even though $H\bigl(k\bigr)$
has no further conditions on $S^{1}\times\bigl(0,\,\pi\bigr)$, on
the boundary circles, $S^{1}\times\Set{0}$ and $S^{1}\times\Set{\pi}$,
however, we must obey \eqref{eq:TRI for Bloch Decomposed}, because
the partners $k$ and $-k$ do both belong to the EBZ for these circles.

To classify the maps from the cylinder $S^{1}\times\bigl[0,\,\pi\bigr]\to S^{4}$
with special boundary conditions, we start by classifying maps from
the boundary circles $S^{1}\to S^{4}$ with special boundary conditions.
For definiteness pick the lower boundary circle $S^{1}\times\bigl\{ 0\bigr\} $.
Then we write for breviy $d\bigl(k_{1},\,0\bigr)$ as $d_{1}\bigl(k_{1}\bigr)$.
If there were no boundary conditions, this classification would be
given by $\pi_{1}\bigl(S^{4}\bigr)$ which is just $\bigl\{ 1\bigr\} $
and so there is just one trivial class in this case. Otherwise, the
special conditions force that $d_{i}\bigl(0\bigr)=d_{i}\bigl(\pi\bigr)=0$
for all $i\in A\backslash TREI$ and otherwise they force that the
loop $S^{1}\to S^{4}$ is symmetric: 
\[
d_{i}\bigl(-k_{1}\bigr)=\begin{cases}
d_{i}\bigl(k_{1}\bigr) & i\in TREI\\
-d_{i}\bigl(k_{1}\bigr) & i\notin TREI
\end{cases}
\]
so that it suffices to specify what happens just from $k_{1}=0$ until
$k_{1}=\pi$. Thus, the class of loops $S^{1}\to S^{4}$ obeying the
special boundary conditions is the class of \emph{paths} $\bigl[0,\,\pi\bigr]\to S^{4}$
which end and start on 
\[
\Set{d\in S^{4}|d_{i}=0\forall i\in A\backslash TREI}=:\hat{S}
\]
In general the possible classes of these maps are all nullhomotopic
if $\hat{S}$ is a path connected subset of $S^{4}$: given a mapping
$d:S^{1}\to S^{4}$ that is not constant, we may find a path $\gamma:\bigl[0,\,1\bigr]\to\hat{S}$
connecting $\gamma\bigl(0\bigr)=d\bigl(0\bigr)$ and $\gamma\bigl(1\bigr)=d\bigl(\pi\bigr)$
and then define a homotopy between a path $d:\bigl[0,\,\pi\bigr]\to S^{4}$
and a loop $\tilde{d}:\bigl[0,\,\pi\bigr]\to S^{4}$ which follows
$\gamma$, and so $\tilde{d}\bigl(\pi\bigr)=\tilde{d}\bigl(0\bigr)$.
We may then concatenate this homotopy with a homotopy that shrinks
the loop $\tilde{d}:\bigl[0,\,\pi\bigr]\to S^{4}$ to a point, and
thus $d:S^{1}\to S^{4}$ is nullhomotopic. If, however, $\hat{S}$
is not path-connected, then we see that the classes of paths $d:\bigl[0,\,\pi\bigr]\to S^{4}$
are organized by the path-connected components of $\hat{S}$.
\begin{claim*}
$\hat{S}$ is path-connected if $\bigl|A\backslash TREI\bigr|\leq1$.
$\hat{S}$ has two path-connected components if $\bigl|A\backslash TRIE\bigr|\in\Set{2,\,3,\,4}$.
There are no other possibilities.
\begin{proof}
As the largest value of $\bigl|A\bigr|$ is five we see that the
possibilities are $\bigl|A\backslash TREI\bigr|\in\Set{0,\,1,\,2,\,3,\,4,\,5}$.
First note that if $\bigl|A\backslash TREI\bigr|=5$ then $d_{i}\bigl(0\bigr)=0$
for all $i\in A$ and so, in particular, $d\bigl(0\bigr)\notin S^{4}$.
As a result, it is not possible that $\bigl|A\backslash TREI\bigr|=5$.

If $\bigl|A\backslash TREI\bigr|=0$, then $\hat{S}\equiv\Set{d\in S^{4}|d_{i}=0\forall i\in\underbrace{A\backslash TREI}_{\varnothing}}=S^{4}$
which is path-connected.

If $\bigl|A\backslash TREI\bigr|=1$, then $\hat{S}$ is a sphere
of lower dimension 
\[
\hat{S}\equiv\Set{d\in S^{4}|d_{i}=0\forall i\in A\backslash TREI}\cong S^{3}
\]
 which is path-connected.

If $\bigl|A\backslash TREI\bigr|=4$, then since on $\hat{S}$, $d_{i}=0\forall i\in A\backslash TREI$,
there is only one index left which is not zero, call it $e$: $\Set{e}=A\cap TREI$,
and since all other indices are zero, that one index must be $\pm1$,
as we must have at all times $\norm{d}=1$. As a result, we have only
two points $\hat{S}=\Set{d_{N},\,d_{S}}$ where $\bigl(d_{N}\bigr)_{e}=1$,
$\bigl(d_{S}\bigr)_{e}=-1$. The two-point-set has two path-connected
components.

If $\bigl|A\backslash TREI\bigr|\in\Set{2,\,3}$, then it \emph{turns
out} that if we make the additional requirement that the set $\Set{\Gamma_{i}}_{i\in A}$
anti-commutes, then we can find either two or three odd gamma-matrices
which anti-commute, and additionally only \emph{one} even gamma matrix
which also anti-commutes with the other odd ones. To reiterate, in
this case, it turns out that $A\cap TREI=\Set{e}$ just as in the
case $\bigl|A\backslash TREI\bigr|=4$ and so we again have only
two points $\hat{S}=\Set{d_{N},\,d_{S}}$ where $\bigl(d_{N}\bigr)_{e}=1$,
$\bigl(d_{S}\bigr)_{e}=-1$.

In particular, our analysis is valid even when $\bigl|A\bigr|<5$.
\end{proof}
\end{claim*}
As a result of the above claim, we only need to consider the case
where $\hat{S}$ has two path-connected components, which we call
$\hat{S}_{N}$ and $\hat{S}_{S}$. Then $A\cap TREI=\Set{e}$.
\begin{claim*}
$\prod_{k\in TRIM}sgn\bigl(d_{e}\bigl(k\bigr)\bigr)=-1$ iff $d:S^{1}\times\bigl[0,\,\pi\bigr]\to S^{4}$
is not nullhomotopic, and $\prod_{k\in TRIM}sgn\bigl(d_{e}\bigl(k\bigr)\bigr)=+1$
iff $d$ is nullhomotopic.
\begin{proof}
For $d:S^{1}\times\bigl\{ 0\bigr\} \to S^{4}$ (the lower boundary
circle) there are in general four possibilities: 
\begin{itemize}
\item (1) NN: $d\bigl(0,\,0\bigr)\in\hat{S}_{N}$ and $d\bigl(\pi,\,0\bigr)\in\hat{S}_{N}$.
(2) SS: $d\bigl(0,\,0\bigr)\in\hat{S}_{S}$ and $d\bigl(\pi,\,0\bigr)\in\hat{S}_{S}$.
(3) NS: $d\bigl(0,\,0\bigr)\in\hat{S}_{N}$ and $d\bigl(\pi,\,0\bigr)\in\hat{S}_{S}$.
(4) SN: $d\bigl(0,\,0\bigr)\in\hat{S}_{S}$ and $d\bigl(\pi,\,0\bigr)\in\hat{S}_{N}$.
\end{itemize}
and $d:S^{1}\times\bigl\{ \pi\bigr\} \to S^{4}$ (the upper circle)
is classified exactly the same. To classify the full maps $d:S^{1}\times\bigl[0,\,\pi\bigr]\to S^{4}$,
observe that what happens in the interior of the cylinder ($S^{1}\times\bigl(0,\,\pi\bigr)$)
is completely unconstrained and so it is only the two loops on the
boundaries ($d:S^{1}\times\bigl\{ 0\bigr\} \to S^{4}$ and $d:S^{1}\times\bigl\{ \pi\bigr\} \to S^{4}$)
which determine the class of the full map $d:S^{1}\times\bigl[0,\,\pi\bigr]\to S^{4}$. 

Next, observe that we may adiabatically rotate $S^{4}$ so as to exchange
$\hat{S}_{N}\leftrightarrow\hat{S}_{S}$. This can be done independently
$\forall k_{2}\in S^{1}$. As a result we really only have four classes
for the whole cylinder map: 
\begin{itemize}
\item (1) NN at $k_{2}=0$, NN at $k_{2}=\pi$. (2) NN at $k_{2}=0$, NS
at $k_{2}=\pi$. (3) NS at $k_{2}=0$, NN at $k_{2}=\pi$. (4) NS
at $k_{2}=0$, NS at $k_{2}=\pi$.
\end{itemize}
We can write this in a more suggestive form, which codifies the geometry
of $\mathbb{T}^{2}$ by\\
 $\begin{pmatrix}k_{2}=\pi,k_{1}=0 & k_{2}=\pi,k_{1}=\pi\\
k_{2}=0,k_{1}=0 & k_{2}=0,k_{1}=\pi
\end{pmatrix}$: 
\begin{itemize}
\item (1) $\begin{pmatrix}N & N\\
N & N
\end{pmatrix}$. (2) $\begin{pmatrix}N & S\\
N & N
\end{pmatrix}$. (3) $\begin{pmatrix}N & N\\
N & S
\end{pmatrix}$. (4) $\begin{pmatrix}N & S\\
N & S
\end{pmatrix}$.
\end{itemize}
The next freedom we can exploit is to exchange $k_{1}\leftrightarrow k_{2}$
so that a map like $\begin{pmatrix}N & S\\
N & S
\end{pmatrix}$ becomes $\begin{pmatrix}S & S\\
N & N
\end{pmatrix}$, after which we may compose another switch on $S^{4}$ of $\hat{S}_{N}\leftrightarrow\hat{S}_{S}$
to obtain in total: 
\[
\begin{pmatrix}N & S\\
N & S
\end{pmatrix}\mapsto\begin{pmatrix}S & S\\
N & N
\end{pmatrix}\mapsto\begin{pmatrix}N & N\\
N & N
\end{pmatrix}
\]
As a result, we see that there really are only two classes of maps,
indexed by the number of $S$ appearing on $k\in TRIM$: One $S$
means the map is not null homotopic, and no $S$ means the map is
null homotopic. This statement may be encoded in the following expression:
\[
\begin{cases}
\prod_{k\in TRIM}sgn\bigl(d_{e}\bigl(k\bigr)\bigr)=-1 & \mbox{non-nullhomotopic}\\
\prod_{k\in TRIM}sgn\bigl(d_{e}\bigl(k\bigr)\bigr)=+1 & \mbox{null-homotopic}
\end{cases}
\]
\end{proof}
\end{claim*}
We thus have a new meaning for the Kane-Mele index, inspired by \cite{Moore_Balents_2007_PhysRevB.75.121306}.
\end{proof}
\end{rem}

\subsection{Bulk-Edge Correspondence Proof for the Case of a Dirac Hamiltonian\label{sec:Topological Insulators Proof for Dirac Hamiltonians}}

We assume the same assumptions of \cref{claim:Special Formula For Kane-Mele index for Dirac Hamiltonians}.
In particular, we may use \cref{subsec:Dirac-Hamiltonians}. We deal
with the case where $A\cap TREI=\Set{e}$. Otherwise, $H\bigl(k\bigr)$
is nullhomotopic, in which case its edge index is easily zero because
the ellipse-point on $S^{4}$ would never contain the origin and so
for constant $H\bigl(k\bigr)$ there are never zero energy edge modes.

Following \cref{claim:Special Formula For Kane-Mele index for Dirac Hamiltonians}
we have that \cref{eq:Bulk-Z2_Invariant_Definition} reduces to 
\begin{eqnarray*}
\bigl(-1\bigr)^{\mathcal{I}_{KM}\bigl(H\bigr)} & = & sgn\bigl(\bigl.M\bigl(k_{2}\bigr)\bigr|_{k_{2}=0}\bigr)sgn\bigl(\bigl.M\bigl(k_{2}\bigr)\bigr|_{k_{2}=\pi}\bigr)
\end{eqnarray*}
where we have defined $M\bigl(k_{2}\bigr):=d_{e}\bigl(\begin{pmatrix}0\\
k_{2}
\end{pmatrix}\bigr)d_{e}\bigl(\begin{pmatrix}\pi\\
k_{2}
\end{pmatrix}\bigr)$.

So as we go with $k_{2}$ from $0$ to $\pi$, $\mathcal{I}_{KM}\bigl(H\bigr)=0$
if $M\bigl(k_{2}\bigr)$ changes sign an \emph{even} number of times,
whereas $\mathcal{I}_{KM}\bigl(H\bigr)=1$ if $M\bigl(k_{2}\bigr)$
changes sign an \emph{odd} number of times. $M\bigl(k_{2}\bigr)$
changes sign when it is zero, and thus, we are looking for the parity
of the number of zeros of the function $M$ on the domain $k_{2}\in\bigl[0,\,\pi\bigr]$.
In the following we use the notation of \cref{subsec:Dirac-Hamiltonians}:
\begin{claim}
We may assume that $H\bigl(k\bigr)$ has been adiabatically changed
(without closing the gap) that $b^{r}$ lies along the $\hat{e}_{e}$
(recall $e=A\cap TREI$) direction, $b^{r}\perp b^{i}$, and that
$b^{0}\cdot b^{i}=0$ for all $k$.
\begin{proof}
We assume that $\hat{e}^{r}$ lies along the $\hat{e}_{e}$ direction
in gamma-space. This should always be possible to achieve via adiabatic
continuous rotations in gamma-space, which is isomorphic to $\mathbb{R}^{5}$
at each $k_{2}$. Observe that these rotations are continuous in $k_{2}$
(because $b\bigl(k_{2}\bigr)$ is continuous in $k_{2}$), and further
more, this is possible to achieve for each $k_{2}$ adiabatically
because the band gap never closes during these rotations, as $E=\pm\norm{d\bigl(k\bigr)}$
shows ($SO\bigl(5\bigr)$ rotations should not affect the energy
bands). The change to make $b^{r}\perp b^{i}$ should also be possible
without closing the gap. It amounts to shrinking the $\hat{e}^{r}$
component of $b^{i}$ to zero. There should be no obstruction to shrink
the $\hat{e}^{i}$ component of $b^{0}$ to zero, even when $b^{0}\cdot\hat{e}^{r}=0$,
because we may always keep a non-zero $b^{0\perp}$ to keep the gap
open. Notice that we do all these changes while keeping each component
$d_{j}$ even or odd respectively in $k$. 

First, to make sure that $b^{r}\perp b^{i}$, examine $d_{e}\bigl(k\bigr)$
($b^{r}$ is already along $\hat{e}_{e}$): 
\[
d_{e}\bigl(k\bigr)\equiv b_{e}^{0}\bigl(k_{2}\bigr)+2b_{e}^{r}\bigl(k_{2}\bigr)\cos\bigl(k_{1}\bigr)+2b_{e}^{i}\bigl(k_{2}\bigr)\sin\bigl(k_{1}\bigr)
\]
so that making sure that $b^{r}\perp b^{i}$ means shrinking $b_{e}^{i}\bigl(k_{2}\bigr)\to0$.
As we do this, $d_{e}\bigl(k\bigr)$ stays even in $k$. Next, we
want to make sure that $b^{0}\cdot b^{i}=0$. We have 
\[
d_{i}\bigl(k\bigr)=b_{i}^{0}\bigl(k_{2}\bigr)+2b_{i}^{i}\bigl(k_{2}\bigr)\sin\bigl(k_{1}\bigr)
\]
where the subscript $i$ denotes the $\hat{e}_{i}$ component. Again,
we may shrink $b_{i}^{0}\bigl(k_{2}\bigr)\to0$ while keeping $d_{i}\bigl(k\bigr)$
odd in $k$. 

In conclusion, along the homotopy to our desired $H\bigl(k\bigr)$,
we keep time-reversal invariance \emph{and }the gap.
\end{proof}

\end{claim}

\begin{claim}
$M\bigl(k_{2}\bigr)$ changes sign exactly at those points $k_{2}\in\bigl[0,\,\pi\bigr)$
where there is an edge state incipient out of or into the bulk. Thus
\[
\mathcal{I}_{KM}\bigl(H\bigr)=\mathcal{I}_{KM}^{\sharp}\bigl(H^{\sharp}\bigr)
\]
\begin{proof}
\emph{$ $}
\begin{itemize}
\item Our model for $d_{i,\,j}\bigl(k\bigr)$ is, as given by \cref{def:Dirac-Hamiltonians}:
\[
d_{i,\,j}\bigl(k\bigr)=b_{i,\,j}^{0}\bigl(k_{2}\bigr)+2b_{i,\,j}^{r}\bigl(k_{2}\bigr)\cos\bigl(k_{1}\bigr)+2b_{i,\,j}^{i}\bigl(k_{2}\bigr)\sin\bigl(k_{1}\bigr)
\]
\item Then the ellipse lives on the plane defined by $\hat{e}^{i}$ and
$\hat{e}^{r}$.
\item We know that there are exactly two edge states in the gap via \cref{claim:dirac_number_of_edge_states},
and that they are at the energies $\pm\norm{b^{0\perp}}$. Thus, if
an edge state is incipient at the lower band for some $k_{2}$, there
is another state simultaneously incipient at the upper band. It is
therefore not important to make sure we deal only with incipience
at the lower band or upper band, as those points give exactly the
same count.
\item We know that there is an edge state at a particular $k_{2}$ when
the ellipse defined by 
\[
b^{0\parallel}\bigl(k_{2}\bigr)+2b^{r}\bigl(k_{2}\bigr)\cos\bigl(k_{1}\bigr)+2b^{i}\bigl(k_{2}\bigr)\sin\bigl(k_{1}\bigr)
\]
includes the origin inside of it. Thus, an edge state is exactly incipient
when $b^{0\parallel}\bigl(k_{2}\bigr)$ lies on the ellipse defined
by $2b^{r}\bigl(k_{2}\bigr)\cos\bigl(k_{1}\bigr)+2b^{i}\bigl(k_{2}\bigr)\sin\bigl(k_{1}\bigr)$,
that is, $b^{0\parallel}\bigl(k_{2}\bigr)=2b^{r}\bigl(k_{2}\bigr)\cos\bigl(k_{1}\bigr)+2b^{i}\bigl(k_{2}\bigr)\sin\bigl(k_{1}\bigr)$
for $k_{1}$ defined by the orientation of $b^{0\parallel}\bigl(k_{2}\bigr)$:
$\cos\bigl(k_{1}\bigr):=\underbrace{\frac{b^{0\parallel}}{\norm{b^{0\parallel}}}\hat{e}^{r}}_{1}$
and $\sin\bigl(k_{1}\bigr):=\underbrace{\frac{b^{0\parallel}}{\norm{b^{0\parallel}}}\hat{e}^{i}}_{0}$.
\item Thus by components we have: 
\[
\begin{cases}
\underbrace{b^{0\parallel}\bigl(k_{2}\bigr)\cdot\hat{e}^{i}}_{0} & =\underbrace{2b^{r}\bigl(k_{2}\bigr)\cdot\hat{e}^{i}}_{0}\cos\bigl(k_{1}\bigr)+2b^{i}\bigl(k_{2}\bigr)\cdot\hat{e}^{i}\underbrace{\sin\bigl(k_{1}\bigr)}_{0}\\
\underbrace{b^{0\parallel}\bigl(k_{2}\bigr)\cdot\hat{e}^{r}}_{\bigl(b^{0\parallel}\bigl(k_{2}\bigr)\bigr)_{e}} & =\underbrace{2b^{r}\bigl(k_{2}\bigr)\cdot\hat{e}^{r}}_{\bigl(2b^{r}\bigl(k_{2}\bigr)\bigr)_{e}}\underbrace{\cos\bigl(k_{1}\bigr)}_{1}+2\underbrace{b^{i}\bigl(k_{2}\bigr)\cdot\hat{e}^{r}}_{0}\sin\bigl(k_{1}\bigr)
\end{cases}
\]
\item From the second equation we have $b^{0\parallel}\bigl(k_{2}\bigr)\cdot\hat{e}^{r}=2b^{r}\bigl(k_{2}\bigr)\cdot\hat{e}^{r}\cos\bigl(k_{1}\bigr)$,
but $\hat{e}^{r}$ extracts exactly the $\hat{e}_{e}$ component of
the vectors by hypothesis, so that we get: $\bigl(b^{0\parallel}\bigl(k_{2}\bigr)\bigr)_{e}=2\bigl(b^{r}\bigl(k_{2}\bigr)\bigr)_{e}\cos\bigl(k_{1}\bigr)$
which means $\bigl(b^{0}\bigl(k_{2}\bigr)\bigr)_{e}=2\bigl(b^{r}\bigl(k_{2}\bigr)\bigr)_{e}\cos\bigl(k_{1}\bigr)$
as $b^{0\perp}$ doesn't have any $\hat{e}_{e}$ component by its
definition.
\item So we have $\bigl(b^{0}\bigl(k_{2}\bigr)\bigr)_{e}-2\bigl(b^{r}\bigl(k_{2}\bigr)\bigr)_{e}=0$
and so 
\begin{eqnarray*}
M\bigl(k_{2}\bigr) & \equiv & \bigl[\bigl(b^{0}\bigl(k_{2}\bigr)\bigr)_{e}-2\bigl(b^{r}\bigl(k_{2}\bigr)\bigr)_{e}\bigr]\bigl[\bigl(b^{0}\bigl(k_{2}\bigr)\bigr)_{e}+2\bigl(b^{r}\bigl(k_{2}\bigr)\bigr)_{e}\bigr]\\
 & = & 0
\end{eqnarray*}
at this point. The other possibility is that $b^{0\parallel}\bigl(k_{2}\bigr)\cdot\hat{e}^{r}=-\bigl(b^{0\parallel}\bigl(k_{2}\bigr)\bigr)_{e}$
which makes the other term zero. 
\item Either way, $M\bigl(k_{2}\bigr)$ changes sign at that point.
\end{itemize}
\end{proof}
\end{claim}

\subsection{Proof for the General Case}

So far we have shown the correspondence for a subclass of Hamiltonians
which we call ``Dirac''. It turns out that for these simple systems,
we are still able to find a non-zero $\mathbb{Z}_{2}$ invariant and
so in this sense, we have already achieved the goal of examining the
simplest non-trivial system. 

In case $H$ is not of ``Dirac'' form, we have no simple formula
which characterizes the existence of edge states as in \cref{subsec:Dirac-Hamiltonians}.
However, it may be possible to generalize \cref{eq:New Formula for Kane-Mele Index}
in the following sense. A ``Dirac'' Hamiltonian has four bands,
which come in two pairs. The lower pair and upper pair are always
degenerate along $k_{2}$. In contrast, the generic Hamiltonian has
four bands which intersect only on $TRIM$. However, it is always
possible to make a time-reversal invariant homotopy which would squeeze
together the lower pair and the upper pair, so as to ultimately bring
it to the ``Dirac'' form. Once this has been made, \eqref{eq:New Formula for Kane-Mele Index}
may be applied. Thus, future work might examine the prescription of
how to make this time-reversal invariant homotopy, which would prescribe
which indices of the fifteen indices ends up being in $A$, and thus
providing a generalization of \eqref{eq:New Formula for Kane-Mele Index}.

\section{Appendix}

\subsection{Explicit Formulation of IQHE Edge Invariant}

Let $\Set{E_{i}^{\#}\bigl(k_{2}\bigr)}_{i\in I}$ denote the discrete
spectrum of $H^{\#}\bigl(k_{2}\bigr)$. Assuming the following set
is of finite order, we define: 
\begin{equation}
\mathcal{D}:=\Set{\bigl(k_{2},\,i\bigr)\in\bigl[0,\,2\pi\bigr)\times I|E_{i}^{\#}\bigl(k_{2}\bigr)=E_{F}\land\bigl(E_{i}^{\sharp}\bigr)'\bigl(k_{2}\bigr)\neq0}\label{eq:CROSS}
\end{equation}
where $E_{F}\in\mathbb{R}$ is the Fermi energy. We assume further
that $\nexists\bigl(k_{2},\,i\bigr)\in\bigl[0,\,2\pi\bigr)\times I$
such that $\bigl(E_{i}^{\sharp}\bigr)'\bigl(k_{2}\bigr)=0$. Under
these assumptions, $\mathcal{D}$ contains all the points where the
edge energy crosses the Fermi energy.
\begin{defn}
\label{def:The-edge-quantum-Hall-conductanc}(\emph{Edge IQHE Topological
Invariant}) We define $\mathcal{I}_{QH}^{\sharp}\bigl(H^{\sharp}\bigr)$
as the signed cardinality in $\mathcal{D}$, that is: 
\begin{equation}
\mathcal{I}_{QH}^{\sharp}\bigl(H^{\sharp}\bigr):=-\sum_{\bigl(k_{2},\,i\bigr)\in\mathcal{D}}\frac{\bigl(E_{i}^{\sharp}\bigr)'\bigl(k_{2}\bigr)}{\bigl|\bigl(E_{i}^{\sharp}\bigr)'\bigl(k_{2}\bigr)\bigr|}\label{eq:Edge_Hall_Conductance}
\end{equation}
which matches our earlier definition, but merely gives an explicit
formula.
\end{defn}

\subsection{\label{subsec:Edge-Spectrum-via-Complex-Momentum}Edge Spectrum via
Complex Momentum}

We can also describe the edge spectrum it via $H$ directly instead
of $H^{\sharp}$, by employing the Dirichlet boundary conditions on
linear combinations of bulk solutions back in real space. At values
of $k_{1}$ that have a strictly positive imaginary part, such solutions
will decay into the right so that they may honestly be called edge
states. Hence to find the additional edge spectrum that will be added
to the bulk spectrum, we follow the procedure: (1) For fixed $k^{\perp}$,
let $E\bigl(k^{\perp}\bigr)\in\mathbb{R}\backslash\sigma\bigl(H\bigl(q\bigr)\bigr)$
be given. (2) Find two (or more) linearly independent eigenstates
$\psi\bigl(k_{1}^{\bigl(1\bigr)},\,k^{\perp}\bigr)$ and $\psi\bigl(k_{1}^{\bigl(2\bigr)},\,k^{\perp}\bigr)$
of $H\bigl(k_{1}^{\bigl(1\bigr)},\,k^{\perp}\bigr)$ and $H\bigl(k_{1}^{\bigl(2\bigr)},\,k^{\perp}\bigr)$
respectively with the same eigenvalue $E\bigl(k^{\perp}\bigr)$ for
some $k_{1}^{\bigl(i\bigr)}$ with strictly positive imaginary part.
(3) Imposing the boundary condition on a linear combination of these
eigenstates gives an additional equation from which $E\bigl(k^{\perp}\bigr)$
may be extracted. This is the value of the edge state at that value
of $k^{\perp}$, if it exists.

\subsection{\label{subsec:Dirac-Hamiltonians}Dirac Hamiltonians}

In this appendix we recall the results of \cite{Mong_Shivamoggi_PhysRevB.83.125109}.
The purpose of repeating the analysis is mainly to bridge the notation
of \cite{Mong_Shivamoggi_PhysRevB.83.125109} with that of the rest
of the paper as well as for the convenience of the reader.
\begin{defn}
\label{def:Dirac-Hamiltonians}Dirac Hamiltonians are Hamiltonians
(after Bloch reduction, thus specified with $H\bigl(k\bigr)$ for
all $k\in\mathbb{T}^{2}$) given by 
\begin{equation}
H\bigl(k\bigr)=\sum_{j=1}^{m}h_{j}\bigl(k\bigr)\Gamma_{j}\label{eq:Dirac_Bulk_Hamiltonian}
\end{equation}
where $\Set{\Gamma_{j}}_{j=1}^{m}$ is a traceless set of Hermitian
$N\times N$ matrices obeying the Clifford algebra 
\[
\bigl\{ \Gamma_{i},\,\Gamma_{j}\bigr\} =2\delta_{ij}\mathds{1}_{N\times N}
\]
and $h$ is a map $\mathbb{T}^{2}\to\mathbb{R}^{m}$. Note that summation
convention on repeating subscript latin indices (such as $j$) will
be assumed and we also omit $k_{2}$ when it is fixed in what follows.
Finally note $\sigma\bigl(H\bigl(k\bigr)\bigr)=\Set{\pm\norm{h\bigl(k\bigr)}}$
and that the gap condition for $H$ implies $h\bigl(k\bigr)\neq0\forall k\in\mathbb{T}^{2}$.
\end{defn}

\begin{claim}
Due to the nearest-neighbor assumption, $h\bigl(k\bigr)$ is of the
form 
\[
h\bigl(k\bigr)=b^{0}+be^{-ik_{1}}+\overline{b}e^{ik_{1}}
\]
where $b^{0}\in\mathbb{R}^{m}$ is given by the components $\bigl\langle b^{0},\,\hat{e}_{i}\bigr\rangle =\frac{1}{N}Tr\bigl[V\bigl(k_{2}\bigr)\Gamma_{i}\bigr]$
and $b\in\mathbb{C}^{m}$ is given by $\bigl\langle b,\,\hat{e}_{i}\bigr\rangle =\frac{1}{N}Tr\bigl[A\bigl(k_{2}\bigr)\Gamma_{i}\bigr]$.

\end{claim}

\begin{rem}
Observe that for a fixed $k_{2}$ and varying $k_{1}$ we have 
\begin{eqnarray*}
h\bigl(k\bigr) & = & b^{0}\bigl(k_{2}\bigr)+2b^{r}\bigl(k_{2}\bigr)\cos\bigl(k_{1}\bigr)+2b^{i}\bigl(k_{2}\bigr)\sin\bigl(k_{1}\bigr)
\end{eqnarray*}
where $b^{r}$ and $b^{i}$ are the real and imaginary vectors of
$b\in\mathbb{C}^{m}$. So at fixed $k_{2}$, $\bigl.h\bigl(k\bigr)\bigr|_{k_{2}}$
traces an ellipse in $\mathbb{R}^{m}$ as $k_{1}$ is varied on $S^{1}$.
Note that this is a feature of the nearest neighbor approximation.
This ellipse lives on the plane spanned by the two vectors $b^{r}\bigl(k_{2}\bigr)$
and $b^{i}\bigl(k_{2}\bigr)$ (and so in particular at different
values of $k_{2}$ this plane changes, but it is independent of $k_{1}$)
and is offset from the origin by the vector $b^{0}$.
\end{rem}

\begin{defn}
Define the following vectors and matrices, most of which are functions
of $k_{2}$ alone unless otherwise noted. We define the two directions
which $b^{r}$ and $b^{i}$ span: $\hat{e}^{r}:=\frac{b^{r}}{\norm{b^{r}}}$,
$\hat{e}^{i}:=\frac{b^{i}-\bigl\langle b^{i},\,\hat{e}^{r}\bigr\rangle \hat{e}^{r}}{\norm{b^{i}-\bigl\langle b^{i},\,\hat{e}^{r}\bigr\rangle \hat{e}^{r}}}$.
It will also be useful to define $\hat{e}^{v}:=\pm\hat{e}^{i}$ (the
sign is unspecified for now). Next we decompose $b^{0}$ along these
two directions: $b^{0\parallel}:=\bigl\langle b^{0},\,\hat{e}^{r}\bigr\rangle \hat{e}^{r}+\bigl\langle b^{0},\,\hat{e}^{i}\bigr\rangle \hat{e}^{i}$,
$b^{0\perp}:=b^{0}-b^{0\parallel}$ and finally we have the direction
spanned by the ``rest'' of $b^{0}$: $\hat{e}^{\perp}:=\frac{b^{0\perp}}{\norm{b^{0\perp}}}$.
Then we may write the planar part of $h$ as:$h^{\parallel}\bigl(k\bigr):=b^{0\parallel}\bigl(k_{2}\bigr)+b\bigl(k_{2}\bigr)e^{-ik_{1}}+\overline{b\bigl(k_{2}\bigr)}e^{ik_{1}}$.
Finally we also expand the Clifford matrices along those axes: $\Gamma^{\alpha}:=e_{j}^{\alpha}\Gamma_{j}\qquad\forall\alpha\in\Set{r,\,i,\,\perp,\,v}$,,
$\Gamma^{\pm}:=\frac{1}{2}\bigl(\Gamma^{r}\pm i\Gamma^{v}\bigr)$,
and write $h^{\alpha}\bigl(k\bigr):=\bigl\langle h\bigl(k\bigr),\,\hat{e}^{\alpha}\bigr\rangle \qquad\forall\alpha\in\Set{r,\,i,\,\perp,\,v}$,
$h^{\pm}\bigl(k\bigr):=h^{r}\bigl(k\bigr)\mp ih^{v}\bigl(k\bigr)$.
\end{defn}

\begin{rem}
Note that $\hat{e}^{v}$ is chosen so that it is orthogonal to $\hat{e}^{r}$
and $\hat{e}^{\perp}$. The reason we don't simply work with $\hat{e}^{i}$
instead is that we want to work in a generality in which the orientation
of the system $\bigl(\hat{e}^{r},\,\hat{e}^{v},\,\hat{e}^{\perp}\bigr)$
is not yet fully specified. This will be then used later in \cref{claim:sign_of_edge_spectrum_in_dirac_Hamiltonians}.
\end{rem}

\begin{claim}
\label{claim:dirac_write_hamiltonian_in_lightcone_gauge}The Hamiltonian
\cref{eq:Dirac_Bulk_Hamiltonian} may be written as 
\[
H\bigl(k\bigr)=\norm{b^{0\perp}\bigl(k_{2}\bigr)}\Gamma^{\perp}+h^{+}\bigl(k\bigr)\Gamma^{+}+h^{-}\bigl(k\bigr)\Gamma^{-}
\]
where we may use the absolute value due to the choice of $\hat{e}^{\perp}$
which is compatible with $b^{0\perp}$.
\begin{proof}
Note that as $h\bigl(k\bigr)$ is spanned by only three vectors,
we may write it as a sum of these components: 
\begin{eqnarray*}
H\bigl(k\bigr) & = & h^{\perp}\bigl(k\bigr)\Gamma^{\perp}+h^{+}\bigl(k\bigr)\Gamma^{+}+h^{-}\bigl(k\bigr)\Gamma^{-}
\end{eqnarray*}
The last step is to recognize that $h^{\perp}$ in fact does not depend
on $k_{1}$: $h^{\perp}\bigl(k\bigr)=\norm{b^{0\perp}\bigl(k_{2}\bigr)}$
using $\hat{e}^{\perp}\bigl(k_{2}\bigr)\perp\Re\bigl\{ b\bigl(k_{2}\bigr)\bigr\} $
and $\hat{e}^{\perp}\bigl(k_{2}\bigr)\perp\Im\bigl\{ b\bigl(k_{2}\bigr)\bigr\} $.
\end{proof}

\end{claim}

\begin{claim}
The following simple facts are verified easily using the Clifford
algebra properties: $\bigl\{ \Gamma^{\perp},\,\Gamma^{\pm}\bigr\} =0$,
$\bigl\{ \Gamma^{+},\,\Gamma^{-}\bigr\} =\mathds{1}$, $\bigl(\Gamma^{\alpha}\bigr)^{2}=\mathds{1}\forall\alpha\in\Set{r,\,i,\,\perp,\,v}$
and $\bigl(\Gamma^{\pm}\bigr)^{2}=0$.
\end{claim}

\begin{rem}
It is assumed that all maps $\mathbb{T}^{2}\to\mathbb{R}$ introduced
so far (the components of $h$ for instance) can be analytically continued
in such a way that $k_{1}$ takes on complex values: 
\[
\eta\bigl(\lambda,\,k_{2}\bigr):=b^{0}\bigl(k_{2}\bigr)+b\bigl(k_{2}\bigr)\lambda^{-1}+\overline{b}\bigl(k_{2}\bigr)\lambda\qquad\forall\lambda\in\mathbb{C}
\]
when a map is analytically continued we denote it by the corresponding
Greek letter: 
\[
\eta\bigl(\exp\bigl(ik_{1}\bigr),\,k_{2}\bigr)\equiv h\bigl(k\bigr)\qquad\forall k\in\mathbb{T}^{2}
\]
\end{rem}

\begin{claim}
\label{claim:Dirac_Only_one_pair_if_energy_is_not_cool}Let $E\in\mathbb{R}$,
$\eta^{\perp}\in\mathbb{R}$ and $u\in\mathbb{C}^{N}\backslash\Set{0}$
be given. \\
If $E\neq\pm\eta^{\perp}$, then the equation 
\begin{equation}
\bigl(\eta^{\perp}\Gamma^{\perp}+\eta^{+}\Gamma^{+}+\eta^{-}\Gamma^{-}\bigr)u=Eu\label{eq:dirac_lightcone_gauge_eigenvalue_equation}
\end{equation}
where $\bigl(\eta^{+},\,\eta^{-}\bigr)$ are unknown variables is
satisfied for \emph{at most} a single pair $\bigl(\eta^{+},\,\eta^{-}\bigr)$.
\begin{proof}
Applying $\bigl(\eta^{\perp}\Gamma^{\perp}+\eta^{+}\Gamma^{+}+\eta^{-}\Gamma^{-}\bigr)$
on \cref{eq:dirac_lightcone_gauge_eigenvalue_equation} from the left
results in 
\begin{eqnarray}
\bigl(\eta^{\perp}\Gamma^{\perp}+\eta^{+}\Gamma^{+}+\eta^{-}\Gamma^{-}\bigr)^{2}u & = & E^{2}u\label{eq:dirac_eigenvalue_equation_squared}
\end{eqnarray}
yet $\bigl(\eta^{\perp}\Gamma^{\perp}+\eta^{+}\Gamma^{+}+\eta^{-}\Gamma^{-}\bigr)^{2}=\bigl[\bigl(\eta^{\perp}\bigr)^{2}+\eta^{+}\eta^{-}\bigr]\mathds{1}$
by the algebra properties so that \cref{eq:dirac_eigenvalue_equation_squared}
becomes 
\begin{eqnarray*}
\eta^{+}\eta^{-}u & = & \bigl[E^{2}-\bigl(\eta^{\perp}\bigr)^{2}\bigr]u
\end{eqnarray*}
But as $u\neq0$ it follows that 
\begin{eqnarray}
\eta^{+}\eta^{-} & = & E^{2}-\bigl(\eta^{\perp}\bigr)^{2}\label{eq:dirac_eval_eq_implication}
\end{eqnarray}
Now assume that $\exists$ two pairs $\bigl(\eta^{+},\,\eta^{-}\bigr)$
and $\bigl(\tilde{\eta}^{+},\,\tilde{\eta}^{-}\bigr)$ such that
\cref{eq:dirac_lightcone_gauge_eigenvalue_equation} is satisfied (observe
that the assumption that $E^{2}\neq\bigl(\eta^{\perp}\bigr)^{2}$
implies $0\notin\Set{\eta^{+},\,\eta^{-},\,\tilde{\eta}^{+},\,\tilde{\eta}^{-}}$).
Then from \cref{eq:dirac_eval_eq_implication} we have 
\begin{equation}
\eta^{+}\eta^{-}=\tilde{\eta}^{+}\tilde{\eta}^{-}\label{eq:dirac_two_pairs_implication}
\end{equation}
as well as $\begin{cases}
\bigl(\eta^{\perp}\Gamma^{\perp}+\eta^{+}\Gamma^{+}+\eta^{-}\Gamma^{-}\bigr)u & =Eu\\
\bigl(\eta^{\perp}\Gamma^{\perp}+\tilde{\eta}^{+}\Gamma^{+}+\tilde{\eta}^{-}\Gamma^{-}\bigr)u & =Eu
\end{cases}$ directly from \cref{eq:dirac_lightcone_gauge_eigenvalue_equation}.
Taking the difference of these two equations gives 
\begin{equation}
\bigl[\bigl(\eta^{+}-\tilde{\eta}^{+}\bigr)\Gamma^{+}+\bigl(\eta^{-}-\tilde{\eta}^{-}\bigr)\Gamma^{+}\bigr]u=0\label{eq:dirac_difference_of_eigenval_eq_on_two_pairs}
\end{equation}
But observe that $\bigl[\bigl(\eta^{+}-\tilde{\eta}^{+}\bigr)\Gamma^{+}+\bigl(\eta^{-}-\tilde{\eta}^{-}\bigr)\Gamma^{+}\bigr]^{2}=\bigl(\eta^{+}-\tilde{\eta}^{+}\bigr)\bigl(\eta^{-}-\tilde{\eta}^{-}\bigr)\mathds{1}$
so that \cref{eq:dirac_difference_of_eigenval_eq_on_two_pairs} implies
(after acting on it from the left with $\bigl[\bigl(\eta^{+}-\tilde{\eta}^{+}\bigr)\Gamma^{+}+\bigl(\eta^{-}-\tilde{\eta}^{-}\bigr)\Gamma^{+}\bigr]$)
$\bigl(\eta^{+}-\tilde{\eta}^{+}\bigr)\bigl(\eta^{-}-\tilde{\eta}^{-}\bigr)=0$
which in turn implies that $\eta^{+}=\tilde{\eta}^{+}$ \emph{or}
$\eta^{-}=\tilde{\eta}^{-}$. We will show that in fact \emph{both}
must hold.

Assume that $\eta^{+}=\tilde{\eta}^{+}$ holds. Then \cref{eq:dirac_two_pairs_implication}
implies $\eta^{+}\eta^{-}=\eta^{+}\tilde{\eta}^{-}$ and as $\eta^{+}\neq0$
we have that $\eta^{-}=\tilde{\eta}^{-}$. The other way around works
similarly and so we conclude the result.
\end{proof}
\end{claim}

\begin{claim}
\label{claim:dirac_edge_State_energy}If for given $E\in\mathbb{R}$,
$k_{2}\in S^{1}$ and $u\in\mathbb{C^{N}\backslash}\Set{0}$ the equation
\begin{equation}
\bigl(\norm{b^{0\perp}\bigl(k_{2}\bigr)}\Gamma^{\perp}+\eta^{+}\bigl(\lambda,\,k_{2}\bigr)\Gamma^{+}+\eta^{-}\bigl(\lambda,\,k_{2}\bigr)\Gamma^{-}\bigr)u=Eu\label{eq:dirac_edge_eval_eqn}
\end{equation}
has \emph{two} solutions $\bigl(\lambda_{1},\,\lambda_{2}\bigr)$
such that $\bigl|\lambda_{1}\bigr|<1$ and $\bigl|\lambda_{2}\bigr|<1$
then it must be that $E=\pm\norm{b^{0\perp}\bigl(k_{2}\bigr)}$.
\begin{proof}
Assume $E\neq\pm\norm{b^{0\perp}\bigl(k_{2}\bigr)}$. Then using
\cref{claim:Dirac_Only_one_pair_if_energy_is_not_cool} it follows
that there is a \emph{single} pair $\bigl(\eta^{+},\,\eta^{-}\bigr)$
such that \cref{eq:dirac_edge_eval_eqn} holds, which we label as $\bigl(\xi^{+},\,\xi^{-}\bigr)$:
\[
\bigl(\norm{b^{0\perp}\bigl(k_{2}\bigr)}\Gamma^{\perp}+\xi^{+}\Gamma^{+}+\xi^{-}\Gamma^{-}\bigr)u=Eu
\]
and we label $\xi_{j}$ the corresponding vector defined by $\bigl(\xi^{+},\,\xi^{-}\bigr)$:
$\xi_{j}=\norm{b^{0\perp}\bigl(k_{2}\bigr)}\hat{e}_{j}^{\perp}+\xi^{+}\hat{e}^{+}+\xi^{-}\hat{e}^{-}$.
Now we would like to find out what is $\lambda\in\mathbb{C}$ corresponding
to this pair $\bigl(\xi^{+},\,\xi^{-}\bigr)$ and so we have to solve
the following equation for $\lambda$ ($k_{2}$ is fixed and suppressed):
\[
\xi_{j}\stackrel{!}{=}\eta_{j}\equiv b_{j}^{0}+b_{j}\lambda^{-1}+\overline{b_{j}}\lambda
\]
which implies $\overline{b_{j}}\lambda^{2}+\bigl(b_{j}^{0}-\xi_{j}\bigr)\lambda+b_{j}=0$
and now using Vieta's formula (which holds for quadratic equations
over $\mathbb{C}$ as well) we have that 
\[
\lambda_{1}\lambda_{2}=\frac{b_{j}}{\bigl(\overline{b_{j}}\bigr)}
\]
Taking the absolute value of this equation we find that $\bigl|\lambda_{1}\bigr|\bigl|\lambda_{2}\bigr|=1$
which implies that $\bigl|\lambda_{1}\bigr|=\frac{1}{\bigl|\lambda_{2}\bigr|}$.
If $\bigl|\lambda_{1}\bigr|<1$ that means that $\bigl|\lambda_{2}\bigr|>1$
which contradicts the initial hypothesis and likewise for $\bigl|\lambda_{2}\bigr|<1$
we have $\bigl|\lambda_{1}\bigr|>1$, again, a contradiction. So
it must be that $E=\pm\norm{b^{0\perp}\bigl(k_{2}\bigr)}$, as desired. 
\end{proof}
\end{claim}

\begin{claim}
\label{claim:Dirac_Hamiltonians_Edge_Spectrum}The edge system $H^{\sharp}\bigl(k_{2}\bigr)$
has a decaying solution at some $k_{2}$ if and only if the ellipse
traced by $h^{\parallel}\bigl(k\bigr)$ ($k_{2}$ is fixed and $k_{1}$
is the parameter along the ellipse) encloses the origin of $\mathbb{R}^{m}$.
If this condition is met, then the energy of that edge state is $E^{\sharp}\bigl(k_{2}\bigr)=\pm\norm{b^{0\perp}\bigl(k_{2}\bigr)}$.
\begin{proof}
From \cref{subsec:Edge-Spectrum-via-Complex-Momentum}, as the first
step, we are looking for a solution $\psi^{\sharp}\in l^{2}\bigl(\mathbb{Z};\,\mathbb{C}^{N}\bigr)$
to the equations 
\[
b_{j}^{0}\Gamma_{j}\psi_{n}^{\sharp}+b_{j}\Gamma_{j}\psi_{n-1}^{\sharp}+\overline{b_{j}}\Gamma_{j}\psi_{n+1}^{\sharp}=E^{\sharp}\psi_{n}^{\sharp}\quad\forall n\in\mathbb{N}
\]
together with the boundary condition that $\psi_{0}^{\sharp}\stackrel{!}{=}0$.
Make an Ansatz of the form $\psi_{n}^{\sharp}=\sum_{j}u_{j}\lambda_{j}^{n}$
(finite sum) to obtain
\begin{eqnarray*}
\sum_{l}\bigl(b_{j}^{0}\Gamma_{j}u_{l}\lambda_{l}^{n}+b_{j}\Gamma_{j}u_{l}\lambda_{l}^{n-1}+\overline{b_{j}}\Gamma_{j}u_{l}\lambda_{l}^{n+1}\bigr) & = & \sum_{l}E^{\sharp}u_{l}\lambda_{l}^{n}
\end{eqnarray*}
so that (omitting $l$ for brevity, but the following holds for each
$l$): 
\begin{eqnarray}
b_{j}^{0}\Gamma_{j}u\lambda^{n}+b_{j}\Gamma_{j}u\lambda^{n-1}+\overline{b_{j}}\Gamma_{j}u\lambda^{n+1} & = & E^{\sharp}u\lambda^{n}\label{eq:edge_eigenvalue_equation}
\end{eqnarray}
This Ansatz makes sense if $\bigl|\lambda\bigr|<1$ as then our solution
indeed decays into the bulk. This can be thought of as a generalized
Bloch solution with $\lambda=\exp\bigl(ik_{1}\bigr)$ where now $\Im\bigl(k_{1}\bigr)>0$.
From \cref{eq:edge_eigenvalue_equation} we have 
\begin{eqnarray}
\lambda\bigl\{ \bigl[b^{0}+b\lambda^{-1}+\overline{b}\lambda^{1}\bigr]\cdot\Gamma-E^{\sharp}\mathds{1}\bigr\} u & = & 0\label{eq:edge_eigenvalue_equation-1}
\end{eqnarray}
Thus we have: 
\begin{equation}
\lambda\bigl[\eta_{j}\bigl(\lambda\bigr)\Gamma_{j}-E^{\sharp}\mathds{1}\bigr]u=0\label{eq:edge_eigenvalue_equation-2}
\end{equation}
and so using \cref{eq:dirac_lightcone_gauge_eigenvalue_equation} we
have 
\begin{eqnarray}
\lambda\bigl[\norm{b^{0\perp}\bigl(k_{2}\bigr)}\Gamma^{\perp}+\eta^{+}\bigl(\lambda,\,k_{2}\bigr)\Gamma^{+}+\eta^{-}\bigl(\lambda,\,k_{2}\bigr)\Gamma^{-}-E^{\sharp}\mathds{1}\bigr]u & = & 0\label{eq:dirac_eval_eq}
\end{eqnarray}
which implies (using the same procedure as in \cref{claim:Dirac_Only_one_pair_if_energy_is_not_cool})
the equation 
\begin{equation}
\lambda^{2}\bigl[\eta^{+}\bigl(\lambda,\,k_{2}\bigr)\eta^{-}\bigl(\lambda,\,k_{2}\bigr)+\norm{b^{0\perp}\bigl(k_{2}\bigr)}^{2}-\bigl(E^{\sharp}\bigr)^{2}\bigr]=0\label{eq:edge-lambda_equation}
\end{equation}
Note that contrary to how the eigenvalue equation is usually solved
($E^{\sharp}$ would be the unknown), we consider the unknown in \cref{eq:edge-lambda_equation}
to be $\lambda$ while $k_{2}$ and $E^{\sharp}$ are fixed.

The next two claims are easy and their proof is omitted.
\begin{claim*}
$\overline{\eta_{j}\bigl(\lambda\bigr)}=\eta_{j}\bigl(\frac{1}{\overline{\lambda}}\bigr)$.
\end{claim*}
$ $
\begin{claim*}
\label{claim:conjugate_solutions_to_edge_eigenvalue_equation}If $\lambda\in\mathbb{C}\backslash\Set{0}$
is a solution of \cref{eq:edge-lambda_equation} then so is $\frac{1}{\bigl(\overline{\lambda}\bigr)}$.
\end{claim*}
Thus we conclude that for every solution of \cref{eq:edge-lambda_equation}
within the unit circle, $\lambda=Re^{i\varphi}$ with $R<1$, there
is a solution outside the unit circle $\frac{1}{\overline{\lambda}}=\frac{1}{Re^{-i\varphi}}=R^{-1}e^{i\varphi}$
($R^{-1}>1$). As a result, only half the solutions are decaying into
the bulk and other other solutions correspond to a mirrored chain,
defined on $-\mathbb{N}$.
\begin{claim*}
\cref{eq:edge-lambda_equation} is an equation of order $4$ in $\lambda$.
\end{claim*}
\begin{proof}
We have $\lambda^{2}\eta^{+}\eta^{-}=\bigl(b_{j}^{0\parallel}\lambda+b_{j}+\overline{b_{j}}\lambda^{2}\bigr)\bigl(b_{j}^{0\parallel}\lambda+b_{j}+\overline{b_{j}}\lambda^{2}\bigr)$.
The other terms in the equation are all of order $\lambda^{2}$.
\end{proof}
Thus by the fundamental theorem of algebra \cref{eq:edge-lambda_equation}
has $4$ solutions in the complex plane. In light of \cref{claim:conjugate_solutions_to_edge_eigenvalue_equation},
we have at most $2$ solutions within the unit circle. Call these
two solutions $\lambda_{1}$ and $\lambda_{2}$.

So the most general form of the edge wave function which is decaying
is 
\[
\psi_{n}^{\sharp}=\sum_{i=1}^{2}u_{i}\bigl(\lambda_{i}\bigr)^{n}
\]
where $u_{i}$ is a null-vector of the matrix $\lambda_{i}\bigl[\sum_{j}\eta_{j}\bigl(\lambda_{i}\bigr)\Gamma_{j}-E^{\sharp}\mathds{1}\bigr]$. 

Following the next step of \cref{subsec:Edge-Spectrum-via-Complex-Momentum},
we need to employ the boundary condition and so we set $\psi_{0}^{\sharp}\stackrel{!}{=}0$
which implies that $u_{1}=-u_{2}$ and thus $\lambda_{1}\bigl[\sum_{j}\eta_{j}\bigl(\lambda_{1}\bigr)\Gamma_{j}-E^{\sharp}\mathds{1}\bigr]$
and $\lambda_{2}\bigl[\sum_{j}\eta_{j}\bigl(\lambda_{2}\bigr)\Gamma_{j}-E^{\sharp}\mathds{1}\bigr]$
share a null-vector. But that means that for a fixed $E^{\sharp}$
and $k_{2}$, the equation $\eta_{j}\bigl(\lambda\bigr)\Gamma_{j}u_{1}=E^{\sharp}u_{1}$
has two solutions $\lambda_{1}$ and $\lambda_{2}$ within the unit
circle, so that we may use \cref{claim:dirac_edge_State_energy} to
conclude that if an edge state exists, then $E^{\sharp}\bigl(k_{2}\bigr)=\pm\norm{b^{0\perp}\bigl(k_{2}\bigr)}$,
showing the last part of our claim.

Furthermore, we have $\begin{cases}
\bigl[\norm{b^{0\perp}}\Gamma^{\perp}+\eta_{1}^{+}\Gamma^{+}+\eta_{1}^{-}\Gamma^{-}\bigr]u_{1} & =E^{\sharp}u_{1}\\
\bigl[\norm{b^{0\perp}}\Gamma^{\perp}+\eta_{2}^{+}\Gamma^{+}+\eta_{2}^{-}\Gamma^{-}\bigr]u_{1} & =E^{\sharp}u_{1}
\end{cases}$ where we have abbreviated $\eta_{i}^{\pm}\equiv\eta^{\pm}\bigl(\lambda_{i}\bigr)$.
We can now compute the anti-commutator: 
\begin{eqnarray*}
\bigl\{ \norm{b^{0\perp}}\Gamma^{\perp}+\eta_{1}^{+}\Gamma^{+}+\eta_{1}^{-}\Gamma^{-},\,\norm{b^{0\perp}}\Gamma^{\perp}+\eta_{2}^{+}\Gamma^{+}+\eta_{2}^{-}\Gamma^{-}\bigr\}  & = & \bigl[\norm{b^{0\perp}}^{2}+\eta_{2}^{+}\eta_{1}^{-}+\eta_{2}^{-}\eta_{1}^{+}\bigr]\mathds{1}
\end{eqnarray*}
yet we also have $\bigl\{ \norm{b^{0\perp}}\Gamma^{\perp}+\eta_{1}^{+}\Gamma^{+}+\eta_{1}^{-}\Gamma^{-},\,\norm{b^{0\perp}}\Gamma^{\perp}+\eta_{2}^{+}\Gamma^{+}+\eta_{2}^{-}\Gamma^{-}\bigr\} u=\bigl(E^{\sharp}\bigr)^{2}u$
so that we may conclude 
\[
\begin{cases}
\eta^{+}\bigl(\lambda_{1}\bigr)\eta^{-}\bigl(\lambda_{1}\bigr) & =0\\
\eta^{+}\bigl(\lambda_{2}\bigr)\eta^{-}\bigl(\lambda_{2}\bigr) & =0\\
\eta^{+}\bigl(\lambda_{2}\bigr)\eta^{-}\bigl(\lambda_{1}\bigr)+\eta^{-}\bigl(\lambda_{2}\bigr)\eta^{+}\bigl(\lambda_{1}\bigr) & =0
\end{cases}
\]
As a result, it appears that either $\eta^{+}\bigl(\lambda\bigr)$
has the two roots $\lambda_{1}$ and $\lambda_{2}$, or $\eta^{-}\bigl(\lambda\bigr)$
has two roots $\lambda_{1}$ and $\lambda_{2}$. But the third equation
excludes the possibility that $\eta^{+}$ and $\eta^{-}$ each have
only one root $\lambda_{1}$ and $\lambda_{2}$ respectively.

We now proceed to show that the existence of the edge state at $k_{2}$
means the ellipse traced by $\bigl.h^{\parallel}\bigl(k\bigr)\bigr|_{k_{2}}$
encloses the origin of $\mathbb{R}^{m}$: 
\begin{enumerate}
\item The number of zeros minus the number of poles of $\eta^{+}$ within
the unit circle is given by $\frac{1}{2\pi i}\oint_{z\in S^{1}\subset\mathbb{C}}\frac{\eta^{+'}\bigl(z\bigr)}{\eta^{+}\bigl(z\bigr)}\dif{z}$.
\item But $\eta^{+}$ has one pole (at $\lambda=0$), and so, to have two
zeros, we must have $\frac{1}{2\pi i}\oint\frac{\eta^{+'}\bigl(z\bigr)}{\eta^{+}\bigl(z\bigr)}\dif{z}\stackrel{!}{=}1$.
\item But $\frac{1}{2\pi i}\oint\frac{\eta^{+'}\bigl(z\bigr)}{\eta^{+}\bigl(z\bigr)}\dif{z}\stackrel{!}{=}1$
iff $\eta^{+}\bigl(e^{ik}\bigr)$ wraps around the origin counterclockwise
once, for $k\in\bigl[0,\,2\pi\bigr]$.\\
\item If, however, $\eta^{+}\bigl(e^{ik}\bigr)$ wraps around the origin
clockwise, $\frac{1}{2\pi i}\oint\frac{\eta^{+'}\bigl(z\bigr)}{\eta^{+}\bigl(z\bigr)}\dif{z}=-1$
and so the number of zeros is $0$ for $\eta^{+}$ (thus no edge states
``from'' $\eta^{+}$). But then, that means that $\eta^{-}\bigl(e^{ik}\bigr)$
wraps around the origin counterclockwise (because $\eta^{-}$ is the
conjugate of $\eta^{+}$ when evaluated on the unit circle) and so
$\frac{1}{2\pi i}\oint\frac{\eta^{-'}\bigl(z\bigr)}{\eta^{-}\bigl(z\bigr)}\dif{z}=1$
and so $\eta^{-}$ has two zeros, and thus, gives rise to an edge
states.
\item Observe that both $\eta^{\pm}\bigl(e^{ik_{1}}\bigr)$ trace the same
ellipse in $\mathbb{C}$ which $\bigl.h^{\parallel}\bigl(k\bigr)\bigr|_{k_{y}}$
traces in some skewed plane of $\mathbb{R}^{m}$. So that if $\bigl.h^{\parallel}\bigl(k\bigr)\bigr|_{k_{y}}$
wraps around the origin (for fixed $k_{2}$ and varying $k_{1}$)
then either $\eta^{+}\bigl(\lambda\bigr)$ or $\eta^{-}\bigl(\lambda\bigr)$
has two zeros within the unit circle.
\end{enumerate}
\end{proof}
\end{claim}

\begin{claim}
\label{claim:dirac_eta_plus_minus_contains_two_roots_then_eta_minus_plus_are_different}If
$\eta^{+}\bigl(\lambda\bigr)$ has two roots $\lambda_{1}$ and $\lambda_{2}$
within the unit circle, then either $\eta^{-}\bigl(\lambda_{1}\bigr)\neq\eta^{-}\bigl(\lambda_{2}\bigr)$
or the ellipse lies on a straight line. The same holds when $+$ and
$-$ are interchanged.
\begin{proof}
Assume that $\eta^{+}\bigl(\lambda_{1}\bigr)=\eta^{+}\bigl(\lambda_{2}\bigr)=0$.
A simple calculation shows that
\begin{eqnarray*}
\eta^{+}\bigl(\lambda\bigr) & = & b^{0r}-ib^{0v}+\bigl(b^{r}+ib^{ir}+b^{iv}\bigr)\lambda^{-1}+\bigl(b^{r}-ib^{ir}-b^{iv}\bigr)\lambda
\end{eqnarray*}
and so if $\eta^{+}\bigl(\lambda\bigr)$ has two roots $\lambda_{1}$
and $\lambda_{2}$ it follows from Vieta's formula that 
\[
\lambda_{1}\lambda_{2}=\frac{b^{r}+ib^{ir}+b^{iv}}{b^{r}-ib^{ir}-b^{iv}}
\]
Now also compute 
\begin{eqnarray*}
\eta^{-}\bigl(\lambda\bigr) & = & b^{0r}+ib^{0v}+\bigl(b^{r}+ib^{ir}-b^{iv}\bigr)\lambda^{-1}+\bigl(b^{r}-ib^{ir}+b^{iv}\bigr)\lambda
\end{eqnarray*}
and assume that $\eta^{-}\bigl(\lambda_{1}\bigr)=\eta^{-}\bigl(\lambda_{2}\bigr)$,
which implies that 
\begin{eqnarray*}
\bigl(b^{r}-ib^{ir}+b^{iv}\bigr)\bigl(\lambda_{1}-\lambda_{2}\bigr) & = & \bigl(b^{r}+ib^{ir}-b^{iv}\bigr)\frac{\lambda_{1}-\lambda_{2}}{\lambda_{1}\lambda_{2}}\\
\lambda_{1}\lambda_{2} & = & \frac{b^{r}+ib^{ir}-b^{iv}}{b^{r}-ib^{ir}+b^{iv}}
\end{eqnarray*}
Thus we obtain the constraint 
\begin{eqnarray*}
b^{r}b^{iv} & = & 0
\end{eqnarray*}
Which geometrically means that the ellipse reduces to a straight line
(either along $\hat{e}^{r}$ or along $\hat{e}^{v}$).

The case with $+$ and $-$ interchanged gives the same constraint
and thus leads to the same conclusion.
\end{proof}
\end{claim}

\begin{claim}
\label{claim:sign_of_edge_spectrum_in_dirac_Hamiltonians}When $N=2$
then the sign of $E^{\sharp}\bigl(k_{2}\bigr)$ is given by 
\[
E^{\sharp}\bigl(k_{2}\bigr)=\bigl[\bigl(\hat{e}^{r}\times\hat{e}^{i}\bigr)\cdot\hat{e}^{\perp}\bigr]\norm{b^{0\perp}\bigl(k_{2}\bigr)}
\]
\begin{proof}
Let $k_{2}\in S^{1}$ be given. \emph{Assume that for $k_{2}$, the
ellipse does not lie on a straight line. }

First note that we may adiabatically (without closing the gap) apply
a unitary transformation on $H\bigl(k\bigr)$, continuously in $k$,
such that $\hat{e}^{r}=e_{1}$, $\hat{e}^{v}=e_{2}$ and $\hat{e}^{\perp}=e_{3}$.
Rotations will not change the magnitude of the vector $\norm{h\bigl(k\bigr)}$
and so will not close the gap, and clearly rotations are continuous.
This is exactly possible because $\hat{e}^{v}$ has an \emph{unspecified}
sign, and so we can make sure that $\bigl(\hat{e}^{r},\,\hat{e}^{v},\,\hat{e}^{\perp}\bigr)$
has right-handed orientation just as the standard basis. As a result,
we will have $\hat{e}^{i}=\pm e_{2}$ and so 
\[
sgn\bigl(b^{iv}\bigr)=sgn\bigl(\bigl\langle b^{i},\,\hat{e}_{2}\bigr\rangle \bigr)=\pm1
\]
The sign of $b^{iv}$ thus matches the sign of $\bigl[\bigl(\hat{e}^{r}\times\hat{e}^{i}\bigr)\cdot\hat{e}^{\perp}\bigr]$.
So if $\bigl[\bigl(\hat{e}^{r}\times\hat{e}^{i}\bigr)\cdot\hat{e}^{\perp}\bigr]=+1$
then $sign\bigl(b^{iv}\bigr)=+1$ and so $\bigl.h^{-}\bigl(k\bigr)\bigr|_{k_{y}}$
goes counter-clockwise in $\mathbb{C}$ whereas if $\bigl[\bigl(\hat{e}^{r}\times\hat{e}^{i}\bigr)\cdot\hat{e}^{\perp}\bigr]=-1$
then $sign\bigl(b^{iv}\bigr)=-1$ then it is $\bigl.h^{+}\bigl(k\bigr)\bigr|_{k_{y}}$
that goes counter-clockwise in $\mathbb{C}$. This can be seen from
\begin{eqnarray*}
\bigl.h^{\pm}\bigl(k\bigr)\bigr|_{k_{2}} & = & \bigl(b^{0r}\mp ib^{0v}\bigr)+2b^{r}\cos\bigl(k_{1}\bigr)+2\bigl(b^{ir}\mp ib^{iv}\bigr)\sin\bigl(k_{1}\bigr)
\end{eqnarray*}
As we know from \cref{claim:Dirac_Hamiltonians_Edge_Spectrum}, the
one of $h^{+}$ or $h^{-}$ which goes counter-clockwise is the one
of $\eta^{+}$ or $\eta^{-}$ that has the two zeros within the unit
circle (if it contains the origin). In conclusion:
\begin{itemize}
\item $\bigl[\bigl(\hat{e}^{r}\times\hat{e}^{i}\bigr)\cdot\hat{e}^{\perp}\bigr]=+1$
means $\eta^{-}$ is the one that might have two zeros within the
unit circle.
\item $\bigl[\bigl(\hat{e}^{r}\times\hat{e}^{i}\bigr)\cdot\hat{e}^{\perp}\bigr]=-1$
means $\eta^{+}$ is the one that might have two zeros within the
unit circle.
\end{itemize}
Assuming we have these relations, we may work with an explicit form
of the three gamma matrices: $\Gamma^{\perp}=\begin{pmatrix}1 & 0\\
0 & -1
\end{pmatrix}$,$\Gamma^{\perp}-\mathds{1}=\begin{pmatrix}0 & 0\\
0 & -2
\end{pmatrix}$, $\Gamma^{\perp}+\mathds{1}=\begin{pmatrix}2 & 0\\
0 & 0
\end{pmatrix}$, $\Gamma^{r}=\begin{pmatrix}0 & 1\\
1 & 0
\end{pmatrix}$, $\Gamma^{v}=\begin{pmatrix}0 & -i\\
i & 0
\end{pmatrix}$, $\Gamma^{+}=\begin{pmatrix}0 & 1\\
0 & 0
\end{pmatrix}$, $\Gamma^{-}=\begin{pmatrix}0 & 0\\
1 & 0
\end{pmatrix}$. 

Next, divide the analysis into two cases:
\begin{enumerate}
\item Assume that it is $\eta^{+}$ that has the two roots (rather than
$\eta^{-}$) within the unit circle. So in this case $\bigl[\bigl(\hat{e}^{r}\times\hat{e}^{i}\bigr)\cdot\hat{e}^{\perp}\bigr]=-1$.

From \cref{eq:dirac_eval_eq} we have that 
\[
\begin{cases}
\bigl[\norm{b^{0\perp}\bigl(k_{2}\bigr)}\Gamma^{\perp}+\eta^{-}\bigl(\lambda_{1},\,k_{2}\bigr)\Gamma^{-}-E^{\sharp}\mathds{1}\bigr]u & =0\\
\bigl[\norm{b^{0\perp}\bigl(k_{2}\bigr)}\Gamma^{\perp}+\eta^{-}\bigl(\lambda_{2},\,k_{2}\bigr)\Gamma^{-}-E^{\sharp}\mathds{1}\bigr]u & =0
\end{cases}
\]
Now we have two subcases: 
\begin{enumerate}
\item $E^{\sharp}=+\norm{b^{0\perp}\bigl(k_{2}\bigr)}$:

In this case we have 
\[
\begin{cases}
\bigl[\norm{b^{0\perp}\bigl(k_{2}\bigr)}\bigl(\Gamma^{\perp}-\mathds{1}\bigr)+\eta^{-}\bigl(\lambda_{1},\,k_{2}\bigr)\Gamma^{-}\bigr]u & =0\\
\bigl[\norm{b^{0\perp}\bigl(k_{2}\bigr)}\bigl(\Gamma^{\perp}-\mathds{1}\bigr)+\eta^{-}\bigl(\lambda_{2},\,k_{2}\bigr)\Gamma^{-}\bigr]u & =0
\end{cases}
\]
which translates into 
\[
\begin{cases}
\bigl\{ \norm{b^{0\perp}\bigl(k_{2}\bigr)}\begin{pmatrix}0 & 0\\
0 & -2
\end{pmatrix}+\eta^{-}\bigl(\lambda_{1},\,k_{2}\bigr)\begin{pmatrix}0 & 0\\
1 & 0
\end{pmatrix}\bigr\} u & =0\\
\bigl\{ \norm{b^{0\perp}\bigl(k_{2}\bigr)}\begin{pmatrix}0 & 0\\
0 & -2
\end{pmatrix}+\eta^{-}\bigl(\lambda_{2},\,k_{2}\bigr)\begin{pmatrix}0 & 0\\
1 & 0
\end{pmatrix}\bigr\} u & =0
\end{cases}
\]
from which we learn that 
\[
\begin{cases}
\eta^{-}\bigl(\lambda_{1},\,k_{2}\bigr)u^{\bigl(1\bigr)}-2\norm{b^{0\perp}\bigl(k_{2}\bigr)}u^{\bigl(2\bigr)}=0\\
\norm{b^{0\perp}\bigl(k_{2}\bigr)}\eta^{-}\bigl(\lambda_{2},\,k_{2}\bigr)u^{\bigl(1\bigr)}-2\norm{b^{0\perp}\bigl(k_{2}\bigr)}u^{\bigl(2\bigr)}=0
\end{cases}
\]
 or that 
\[
\begin{cases}
u & =\begin{pmatrix}\bigl(\frac{2\norm{b^{0\perp}\bigl(k_{2}\bigr)}}{\eta^{-}\bigl(\lambda_{1},\,k_{2}\bigr)}\bigr)\\
1
\end{pmatrix}\\
u & =\begin{pmatrix}\bigl(\frac{2\norm{b^{0\perp}\bigl(k_{2}\bigr)}}{\eta^{-}\bigl(\lambda_{2},\,k_{2}\bigr)}\bigr)\\
1
\end{pmatrix}
\end{cases}
\]
But then it must be that $\eta^{-}\bigl(\lambda_{1},\,k_{2}\bigr)=\eta^{-}\bigl(\lambda_{2},\,k_{2}\bigr)$,
which, as we learnt in \cref{claim:dirac_eta_plus_minus_contains_two_roots_then_eta_minus_plus_are_different}
is not possible because \emph{by hypothesis} the ellipse is not on
a straight line, so we must conclude this case is not possible.
\item $E^{\sharp}=-\norm{b^{0\perp}\bigl(k_{2}\bigr)}$:

In this case we have 
\[
\begin{cases}
\bigl[\norm{b^{0\perp}\bigl(k_{2}\bigr)}\bigl(\Gamma^{\perp}+\mathds{1}\bigr)+\eta^{-}\bigl(\lambda_{1},\,k_{2}\bigr)\Gamma^{-}\bigr]u & =0\\
\bigl[\norm{b^{0\perp}\bigl(k_{2}\bigr)}\bigl(\Gamma^{\perp}+\mathds{1}\bigr)+\eta^{-}\bigl(\lambda_{1},\,k_{2}\bigr)\Gamma^{-}\bigr]u & =0
\end{cases}
\]
which translates into 
\[
\begin{cases}
\bigl\{ \norm{b^{0\perp}\bigl(k_{2}\bigr)}\begin{pmatrix}2 & 0\\
0 & 0
\end{pmatrix}+\eta^{-}\bigl(\lambda_{1},\,k_{2}\bigr)\begin{pmatrix}0 & 0\\
1 & 0
\end{pmatrix}\bigr\} u & =0\\
\bigl\{ \norm{b^{0\perp}\bigl(k_{2}\bigr)}\begin{pmatrix}2 & 0\\
0 & 0
\end{pmatrix}+\eta^{-}\bigl(\lambda_{2},\,k_{2}\bigr)\begin{pmatrix}0 & 0\\
1 & 0
\end{pmatrix}\bigr\} u & =0
\end{cases}
\]
from which we learn that 
\[
\begin{cases}
\begin{cases}
2\norm{b^{0\perp}\bigl(k_{2}\bigr)}u^{\bigl(1\bigr)} & =0\\
\eta^{-}\bigl(\lambda_{1},\,k_{2}\bigr)u^{\bigl(1\bigr)} & =0
\end{cases}\\
\begin{cases}
2\norm{b^{0\perp}\bigl(k_{2}\bigr)}u^{\bigl(1\bigr)} & =0\\
\eta^{-}\bigl(\lambda_{2},\,k_{2}\bigr)u^{\bigl(1\bigr)} & =0
\end{cases}
\end{cases}
\]
 or that 
\[
u=\begin{pmatrix}0\\
1
\end{pmatrix}
\]
which leads to no contradictions.
\end{enumerate}
We have thus shown that 
\[
\bigl[\bigl(\hat{e}^{r}\times\hat{e}^{i}\bigr)\cdot\hat{e}^{\perp}\bigr]=-1\quad\boxed{\Longrightarrow}\quad E^{\sharp}=-\norm{b^{0\perp}\bigl(k_{2}\bigr)}
\]
\item The other case leads to the complementary conclusion, namely, if $\eta^{-}\bigl(\lambda\bigr)$
has two roots within the unit circle then $\bigl[\bigl(\hat{e}^{r}\times\hat{e}^{i}\bigr)\cdot\hat{e}^{\perp}\bigr]=+1$,
and we will find that to avoid contradictions it must be that $E^{\sharp}=+\norm{b^{0\perp}\bigl(k_{2}\bigr)}$.
\end{enumerate}
Thus when the ellipse does not lie on a straight line, we have proven
the formula 
\[
E^{\sharp}\bigl(k_{2}\bigr)=\bigl[\bigl(\hat{e}^{r}\times\hat{e}^{i}\bigr)\cdot\hat{e}^{\perp}\bigr]\norm{b^{0\perp}\bigl(k_{2}\bigr)}
\]
Now, assuming that $E^{\sharp}\bigl(k_{2}\bigr)$ is continuous,
we can take the limit $\hat{e}^{i}\to\pm\hat{e}^{r}$ (then the ellipse
\emph{is} on a straight line). In this limit, $E^{\sharp}\bigl(k_{2}\bigr)\to0$.
So it must be that when the ellipse lies on a straight line, $E^{\sharp}\bigl(k_{2}\bigr)=0$.
\end{proof}
\end{claim}

\begin{claim}
\label{claim:dirac_number_of_edge_states}If $N=4$ and there is an
edge state at a given $k_{2}$ then there are in fact at least two
edge states corresponding to both $E^{\sharp}\bigl(k_{2}\bigr)=+\norm{b^{0\perp}\bigl(k_{2}\bigr)}$
and $E^{\sharp}\bigl(k_{2}\bigr)=-\norm{b^{0\perp}\bigl(k_{2}\bigr)}$.
\begin{proof}
Without loss of generality, assume that $\eta^{+}$ has two zeros
within the unit circle (the case for $\eta^{-}$ proceeds analogously).
Then the eigenvalue equation, depending on the sign of the energy,
is either: 
\[
\begin{cases}
\bigl[\norm{b^{0\perp}\bigl(k_{2}\bigr)}\bigl(\Gamma^{\perp}-\mathds{1}\bigr)+\eta^{-}\bigl(\lambda_{1},\,k_{2}\bigr)\Gamma^{-}\bigr]u & =0\\
\bigl[\norm{b^{0\perp}\bigl(k_{2}\bigr)}\bigl(\Gamma^{\perp}-\mathds{1}\bigr)+\eta^{-}\bigl(\lambda_{2},\,k_{2}\bigr)\Gamma^{-}\bigr]u & =0
\end{cases}\quad\mbox{for }\quad E^{\sharp}\bigl(k_{2}\bigr)=+\norm{b^{0\perp}\bigl(k_{2}\bigr)}
\]
with the same $u$ for both equations, or 
\[
\begin{cases}
\bigl[\norm{b^{0\perp}\bigl(k_{2}\bigr)}\bigl(\Gamma^{\perp}+\mathds{1}\bigr)+\eta^{-}\bigl(\lambda_{1},\,k_{2}\bigr)\Gamma^{-}\bigr]v & =0\\
\bigl[\norm{b^{0\perp}\bigl(k_{2}\bigr)}\bigl(\Gamma^{\perp}+\mathds{1}\bigr)+\eta^{-}\bigl(\lambda_{2},\,k_{2}\bigr)\Gamma^{-}\bigr]v & =0
\end{cases}\quad\mbox{for }\quad E^{\sharp}\bigl(k_{2}\bigr)=-\norm{b^{0\perp}\bigl(k_{2}\bigr)}
\]
with the same $v$ for both equations. Our goal is to show that both
duos of equations are possible simultaneously with $u$ and $v$ linearly
independent (whereas in \cref{claim:sign_of_edge_spectrum_in_dirac_Hamiltonians}
only one was possible, which allowed us to determine the sign of the
energy of the edge state).

For the case when $N=4$, again we may work without loss of generality
with a particular representation of the Gamma matrices so that: $\Gamma^{\perp}=\sigma_{3}\otimes\sigma_{0}$,
$\Gamma^{r}=\sigma_{1}\otimes\sigma_{0}$, and $\Gamma^{v}=\sigma_{2}\otimes\sigma_{3}$.
Then $\Gamma^{\perp}-\mathds{1}=diag\bigl(0,\,0,\,-2,\,-2\bigr)$,$\Gamma^{\perp}+\mathds{1}=diag\bigl(2,\,2,\,0,\,0\bigr)$
and $\Gamma^{-}=\begin{pmatrix}0 & 0 & 0 & 0\\
0 & 0 & 0 & 1\\
1 & 0 & 0 & 0\\
0 & 0 & 0 & 0
\end{pmatrix}$. 

Then it's easy to verify that both duos of the eigenvalue equations
are possible to satisfy, the first with $u=\begin{pmatrix}0\\
1\\
0\\
0
\end{pmatrix}$ and the second with $v=\begin{pmatrix}0\\
0\\
1\\
0
\end{pmatrix}$, both of which are linearly independent and thus correspond to different
solutions, each of which has an opposite sign of energy. What's more,
neither of these solutions are obstructed by a requirement of the
form $\eta^{-}\bigl(\lambda_{1},\,k_{2}\bigr)=\eta^{-}\bigl(\lambda_{2},\,k_{2}\bigr)$.
\end{proof}
\end{claim}

\textbf{Acknowledgments.} I thank Gian Michele Graf for bringing this
topic to my attention and for interesting discussions.

\bibliographystyle{plain}

\end{document}